\documentclass{ws-jnopm}
\usepackage{graphics}
\usepackage{amsmath}
\usepackage{color}
\usepackage[hypertex]{hyperref}

\begin{document}

\markboth{Rick Lytel, Mark Kuzyk}{Dressed quantum graphs \ldots}

%%%%%%%%%%%%%%%%%%% Publisher's Area please ignore %%%%%%%%%%%%%%%%%%%%%%%
\catchline{}{}{}{}{}
%%%%%%%%%%%%%%%%%%%%%%%%%%%%%%%%%%%%%%%%%%%%%%%%%%%%%%%%%%%%%%%%%%%%%%%%%%

\title{DRESSED QUANTUM GRAPHS WITH OPTICAL NONLINEARITIES APPROACHING THE FUNDAMENTAL LIMIT}

\author{RICK LYTEL and MARK G. KUZYK}

\address{Department of Physics and Astronomy, Washington State University\\ Pullman, Washington 99164-2814\\
rlytel@wsu.edu}

\maketitle

\begin{history}
%\received{(29 August 2013)}
%\revised{(Day Month Year)}
%\accepted{(Day Month Year)}
%\comby{(xxxxxxxxxx)}
\end{history}

\begin{abstract}
{\noindent We dress bare quantum graphs with finite delta function potentials and calculate optical nonlinearities that are found to match the fundamental limits set by potential optimization.  We show that structures whose first hyperpolarizability is near the maximum are well described by only three states, the so-called three-level Ansatz, while structures with the largest second hyperpolarizability require four states.  We analyze a very large set of configurations for graphs with quasi-quadratic energy spectra and show how they exhibit better response than bare graphs through exquisite optimization of the shape of the eigenfunctions enabled by the existence of the finite potentials. We also discover an exception to the universal scaling properties of the three-level model parameters and trace it to the observation that a greater number of levels are required to satisfy the sum rules even when the three-level Ansatz is satisfied and the first hyperpolarizability is at its maximum value, as specified by potential optimization.  This exception in the universal scaling properties of nonlinear optical structures at the limit is traced to the discontinuity in the gradient of the eigenfunctions at the location of the delta potential.  This is the first time that dressed quantum graphs have been devised and solved for their nonlinear response, and it is the first analytical model of a confined dynamic system with a simple potential energy that achieves the fundamental limits.}
\end{abstract}


%%%%%%%%%%%%%%%%%%%%%%%%%%%%%%%%%%%%%%%%%%%%%%%%%%%

\section{Introduction}\label{sec:intro}

Nonlinear optical materials are of great interest for high-speed modulation\cite{chen97.01,woote00.01}, optical bistability and switching\cite{winfu80.01,gibbs84.01}, ultrafast optics\cite{weine11.01}, waveguide switching\cite{vanec91.01}, phase conjugation\cite{yariv77.01,yariv78.01}, photorefraction\cite{khoo87.01}, harmonic generation\cite{maker65.02,bloem68.01,bass69.01}, four-wave mixing\cite{khoo81.01,lytel84.01,lytel86.02,boyd92.01}, self-focusing\cite{khoo82.02}, electro-optics \cite{wayna00.01}, saturable absorption\cite{tutt93.01}, and general light by light control, if their nonlinear response per unit size is large enough for practical applications with common coherent optical sources\cite{horna92.01,kuzyk06.06}.  The figures of merit for the off-resonance, electronic nonlinear optical response are the scale-free, \emph{intrinsic}, fully symmetric third and fourth rank hyperpolarizability tensors $\beta_{ijk}$ and $\gamma_{ijkl}$, respectively, whose magnitudes are limited by\cite{kuzyk00.01,kuzyk00.02}

\begin{equation}\label{sh-betaMax}
\beta_{max} = 3^{1/4} \left(\frac{e\hbar}{\sqrt{m}}\right)^3 \frac{N^{3/2}}{E_{10}^{7/2}}
\end{equation}
and
\begin{equation}\label{sh-gammaMax}
\gamma_{max} = 4 \left(\frac{e\hbar}{\sqrt{m}}\right)^4 \frac{N^{2}}{E_{10}^{5}} ,
\end{equation}
where $E_{10}$ is the energy difference between the first excited state and the ground state, m is the electron mass, e is its charge, and N is the number of electrons.
Figures of merit are then defined by normalizing the first and second hyperpolarizabilities to these maxima, viz.
\begin{equation}\label{IntrinsicBetaGamma}
\beta_{ijk} \rightarrow \frac {\beta_{ijk}} {\beta_{max}} \hspace{2em} \gamma_{ijkl} \rightarrow \frac {\gamma_{ijkl}} {\gamma_{max}} .
\end{equation}
The second hyperpolarizability normalized this way has a largest negative value equal to $-(1/4)$ of the maximum value.

The existence of limits results from the physical constraints that the quantum system, described by a self-adjoint Hamiltonian operator, necessarily satisfies the Thomas-Reiche-Kuhn (TRK) sum rules\cite{wang99.01}.  The first (intrinsic) hyperpolarizability is a sum over states:
\begin{equation}\label{betaInt}
\beta_{ijk} = {\sum_{n,m}}' \beta_{ijk}(n,m) = \left(\frac{3}{4}\right)^{3/4} {\sum_{n,m}}' \frac{\xi_{0n}^{i}\bar{\xi}_{nm}^{j}\xi_{m0}^{k}}{e_n e_m},
\end{equation}
which implicitly defines the eigenfunction density for $\beta_{ijk}$.  In Eq. (\ref{betaInt}), the prime indicates that the sum excludes the ground state, and $\xi_{nm}^{i}$ and $e_n$ are normalized transition moments and energies, defined by
\begin{equation}\label{xNMnorm}
\xi_{nm}^{i} = \frac{r_{nm}^{i}}{r_{01}^{max}}, \qquad e_{n} = \frac{E_{n0}}{E_{10}},
\end{equation}
with $r_{nm}^{i}=\left<n|r^{i}|m\right>$, $E_{n0}=E_{n}-E_{0}$, $r^{i=1}=x$, $r^{i=2}=y$, and $r^{i=3}=z$, and where
\begin{equation}\label{Xmax}
r_{01}^{max} = \left(\frac{\hbar^2}{2 m E_{10}}\right)^{1/2}.
\end{equation}
$r_{01}^{max}$ represents the largest possible transition moment value of $r_{01}$.  According to Eq. (\ref{xNMnorm}), $e_0 = 0$ and $e_1 = 1$. $\beta_{ijk}$ is scale-invariant and can be used to compare molecules of different shapes and sizes.  Similarly, the second intrinsic hyperpolarizability tensor is given by
\begin{eqnarray}\label{gammaInt}
\gamma_{ijkl} &=& {\sum_{n,m,p}}' \gamma_{ijkl}(n,m,p) \\
&=& \frac{1}{4} {\sum_{n,m,p}}' \left(\frac{\xi_{0n}^{i}\bar{\xi}_{nm}^{j}\bar{\xi}_{mp}^{k}\xi_{p0}^{l}-\delta_{mp}\xi_{0n}^{i}\xi_{n0}^{j}\xi_{0m}^{k}\xi_{m0}^{l}}{e_n e_m e_p}\right),\nonumber
\end{eqnarray}
which implicitly defines the eigenfunction density for $\gamma_{ijkl}$.  In principle, systems should exist that allow the hyperpolarizabilities in Eq. (\ref{betaInt}) and Eq. (\ref{gammaInt}) to each achieve a maximum value of unity.

However, it is an empirical fact that all nonlinear optical materials discovered to date fall short of the maximum allowed values, most by over an order of magnitude\cite{kuzyk13.01}.  This is a rather profound observation, inasmuch the search for highly active nonlinear materials is well over 40 years old and continues today.  To explore why this is so, analytical studies of the space of possible sets of transition moments and spectra restricted only by the TRK sum rules were carried out using Monte Carlo methods with the result that the maxima may be achieved, at least numerically\cite{kuzyk08.01,shafe10.01}, though the nature of the Hamiltonian that can produce these is still unknown.

Subsequent studies began with specific potentials and calculated the spectra and states directly in an effort to discover the optimum potential for maximum response.  The shapes of optimized potentials that have the largest hyperpolarizabilities were used to propose modulated conjugation as one new design guideline for making better molecules.\cite{zhou06.01,zhou07.02}  Searches using the concept of modulation of conjugation led to the discovery of molecules with record hyperpolarizabilities\cite{perez07.01,perez09.01} at that time, still well short of the fundamental limits.

However, three major results emerged from the potential optimization studies.  The first is that the theoretical maximum for $\beta$ appears to assume a value that is about $0.7089$, not unity.  For $\gamma$, both extremes appear to be exactly sixty percent of the theoretical limits, viz, $-.15 \leq\gamma\leq 0.6$.  Recent examination of the origin of the limits has suggested that the true limits are indeed those from potential optimization and not those determined solely by the sum rules \cite{shafe13.01}.  Corroborating results have also been provided \cite{ather12.01}, but have shown that optimization poorly determines the potential.

The second major result is that all systems known to date satisfy a set of empirical observations known collectively as the three-level Ansatz (TLA), whereby a structure that appears to be optimized for achieving its best $\beta$ has spectra and states such that the contributions to $\beta$ arise mainly from three states.  This remarkable observation appears to be universal near the optimum value of $\beta$ for any system, in spite of the fact that many of the transition moments among higher states may be comparable or larger than those of the three lower states (and thus satisfying the TRK sum rules). Moreover, the three level model parameters $X\equiv x_{01}/x_{01}^{max}$ and $E\equiv E_{10}/E_{20}$ appear to scale to universal values $X \sim 0.79$ and $E \sim 0.49$, independent of any structural features, when the hyperpolarizability scales to its optimum value.

The final result is that the optimum potentials are those producing an energy spectrum that scales quadratically or faster with eigenenergy.  This observation may explain why most molecular structures that are sampled off-resonance have nonlinearities that are well below the maximum.  Custom design of artificial materials may produce structures with the desired potentials.  Lower dimensional systems would be expected to exhibit additional enhancements due to confinement effects, as well as the change in the scaling properties of the potential. It is also desirable to examine systems with a quasi-quadratic energy spectrum but with sufficient topological variability to permit fine-tuning of the energy differences of the first few eigenstates, as well as control over the shape of the eigenstates in order to maximize the specific transition moments contributing the most to the hyperpolarizability near its optimum value.  One dimensional quantum wells have too regular a spectrum to meet these requirements, but quasi-one dimensional quantum graph models often have regular root boundaries (points between which one and only one eigenenergy can be found) at locations resembling the spectra of a one-dimensional well but have energy spectra that may be varied within a pair of root boundaries in such a manner that energy differences and ratios may be custom-tailored by altering the topology of the graph.

For these reasons, we recently launched a study of the nonlinear optical properties of quasi-one dimensional quantum structures modeled by quantum graphs\cite{shafe12.01,lytel13.01,lytel13.02}.  A quantum graph (QG) is a general confinement model for quasi-one dimensional electron dynamics.  The objectives of the exploration are (1) to determine the exact dependence of the nonlinear optical response on the geometry in each of a class of graph topologies in order to understand the configurations capable of producing eigenvalues and transition moments that yield the largest first and second hyperpolarizabilities, (2) to identify universal properties of the spectra and states as the nonlinearities approach the fundamental limits, and (3) to explore the vast space of physical solutions to any Hamiltonian defined on the graph that are consistent with the Thomas-Reiche-Kuhn (TRK) sum rules and general principles of quantum mechanics.

The generalized QG model of an $N$ electron structure constrains dynamics to the edges of a metric graph.  Our prior studies show that the one-electron version of the generalized quantum graph model (hereafter referred to as the \emph{elementary QG}) exhibits the scaling behavior found in actual nonlinear optical systems.  At the same time, multi-electron systems with arbitrary interactions yield the same universal results as one-electron systems\cite{watki11.01}, suggesting that the single electron models are suitable vehicles for studying the relationships of graph geometry and topology to the scaling properties of the nonlinear optical tensors.  The single electron quantum graph is a well-studied, exactly solvable model of quantum chaos\cite{kotto97.01,kotto99.02,blume02.01,blume02.02,dabag04.01,dabag02.01,dabag03.01,dabag07.01}.  With this in mind, we initiated our studies of the elementary QG model for nonlinear optics by focusing first on undressed edges and calculated the off-resonance first ($\beta_{ijk}$) and second ($\gamma_{ijkl}$) hyperpolarizability tensors (normalized to their maximum values) of elementary graphical structures, such as wires, closed loops, and star vertices\cite{shafe12.01,lytel12.01} and to investigate the relationship between the topology and geometry of a graph and its nonlinear optical response through its hyperpolarizability tensors\cite{lytel13.01}.  We also developed computational methods using graphical motifs in order to solve any graph and applied the method to topologies with several sets of star vertices\cite{lytel13.02}.  With undressed edges, we obtained nonlinear responses for both $\beta$ and $\gamma$ approaching eighty percent of the fundamental limits empirically determined from potential optimization studies\cite{kuzyk08.01,zhou07.02}.

The present work addresses graphs with dressed edges and discovers for the first time topologies achieving the limits obtained using 1D potential optimization.  The quantum wires are dressed with finite delta potentials located anywhere on the wires. These structures are commonly called compressed delta atoms with one delta potential\cite{blume06.01}, and we call them compressed delta molecules with more than one delta potential.  Since these are quasi-1D quantum graphs, we can allow nonzero angles between the edges in order to vary graph geometry, if desired.  We also dress the two star graph motifs\cite{lytel13.02} (the three-star and the lollipop) with finite delta function potentials located at vertices.  The presence of a delta function on a wire or at a vertex creates a source or sink for flux in the graph and provides a means for sensitive control of the nonlinearity by altering the matching of edge functions in such a way that optimum eigenfunction distribution may be achieved.  The models simulate the presence of defects on intersections of quantum wires and show that the presence of such defects may provide significant enhancement of the nonlinearity.  We are able to obtain nearly ninety percent of the potential-optimized upper bound of the first and second hyperpolarizabilities with a single, finite delta potential at the central vertex of a star or lollipop graph, and one hundred percent for the compressed delta atoms and molecules.

This paper is presented as follows.  Section \ref{sec:standardGraphModel} begins with a concise summary of the standard method we developed for calculating the hyperpolarizability tensors of a large ensemble of topologically equivalent graphs whose geometry is varied in a Monte Carlo algorithm\cite{shafe12.01,lytel12.01,lytel13.01,lytel13.02}.  We then invoke the motif method\cite{lytel13.02} to determine the characteristic function of the compressed delta atom in Section \ref{sec:deltaWires}, as preparation for all subsequent work.  Section \ref{sec:deltaAtom} studies the basic quantum wire with a finite delta function.  Section \ref{sec:delta2Atom} places an additional delta function on the wire.  Section \ref{sec:delta3Atom} adds yet another delta function to the wire and shows that the resulting topology hits the maximum possible nonlinearity predicted by the theory of fundamental limits of nonlinear optical materials.  Sections \ref{sec:deltaStar} and \ref{sec:deltaLollipop} solve the two star motif graphs with a central delta potential, show how the dressed graphs achieve larger nonlinearities than the bare graphs, and show how their model parameters scale toward universal values as the geometries of the graphs achieve the form that maximizes the hyperpolarizabilities.  Section \ref{sec:Conclusions} summarizes the key results of this paper and places them within the framework of our prior work on the elementary quantum graph model.

The configurations studied in this paper are mainly linear along the x-direction of the lab frame.  However, the quantum graphs are inherently two-dimensional objects, being comprised of a longitudinal dimension along the edges and a transverse dimension perpendicular to the edges\cite{shafe11.02}.  In the limit of vanishing transverse dimension, the hyperpolarizabilities depend only on the electron motion in the longitudinal direction.  However, the integrity of the model, including its unitarity, require transverse contributions to satisfy the TRK sum rules.  This observation enables sensitive sum rule tests to verify that the calculated eigenstates of every graph are complete in Hilbert space, and that the energy spectra are correct\cite{shafe12.01}.

\section{Calculation of optical nonlinearities of quantum graphs}\label{sec:standardGraphModel}
\begin{figure}\center
\includegraphics[width=4in]{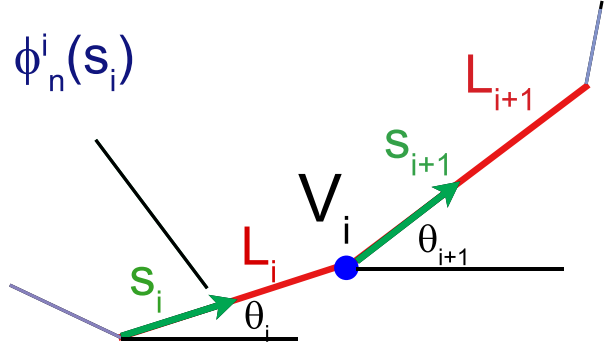}
\caption{Quantum graph, showing the notation for the edges and vertices.  Edge functions are continuous at internal vertices, and the quantum flux is conserved at these vertices.}\label{fig:NEWESTgraph}
\end{figure}

Our present work invokes our standard model of calculating nonlinearities for quantum graphs.  We have used a general Monte Carlo method that explores a very large configuration space to discover structures with optimum nonlinearities, which has been extensively reviewed in our prior publications for undressed graphs\cite{shafe12.01,lytel12.01,lytel13.01,lytel13.02}.  We start by randomly selecting vertices in 2D space, connecting them with metric edges representing a desired topology (eg, a loop or star), endowing them with single-electron dynamics, and then solving for the exact eigenstates and spectra of the entire graph by first computing the edge states that are eigenfunctions of the Hamiltonian on each edge with the same spectrum as all other edges and then using an appropriate union operation on edge states to create an eigenstate of a Hilbert space that is a direct sum of those on each edge\cite{shafe12.01}:

\begin{equation}\label{eigenFunction}
\psi_n(s)= \cup_{i=1}^{E} \phi_n^{i}(s_{i}) ,
\end{equation}
where the notation is that of Fig. \ref{fig:NEWESTgraph}.  The edge functions take the general form
\begin{equation}\label{edgeFunctions2}
\phi_{n}^{i}(s_{i})=A_{n}^{i}\sin{(k_{n}s_{i}+\eta_{n}^{i})}
\end{equation}
whenever the potential is zero on the edges.  At each vertex, the edge functions match in amplitude, and their derivatives sum to a net flux of zero into or out of a vertex.  Terminated ends are modeled as infinite potentials, where edge functions vanish.  The collection of boundary conditions, taken together, provide a set of simultaneous equations whose solution demands that the determinant of the coefficients vanishes.  This produces a transcendental characteristic (secular) equation that provides the eigenvalues for the graph.  The coefficients $A_{n}^{i}$ and the phases $\eta_{n}^{i}$ of the edge functions are then extracted and used to construct the eigenstates through the union operation.  The eigenstates are then normalized, and the set of states and energies may be used to compute the nonlinear optical properties of the graph.

%%%%%%%%%%%%%%%%%%%%%%%%%%%%%
There are always enough boundary conditions to solve for the eigenvalues and amplitudes in a graph.  Let the graph have E edges, $V_{E}$ external vertices, $V_{I}$ internal vertices, and $V=V_{E}+V_{I}$ total vertices.  Each eigenfunction has two unknown amplitudes, so there are exactly 2E unknown amplitudes.  We need exactly 2E boundary condition equations in order to solve for the eigenvalues and get the amplitudes, up to a normalization constant.

There are exactly $V_{E}$ amplitude boundary condition equations on terminated ends.  There are also exactly $V_{I}$ internal flux conservation equations on internal edges.  These two sets provide exactly V boundary condition equations.  Next, let $d^{(k)}$ be the degree of each internal vertex, ie, the number of edges connected at the $kth$ internal vertex.  There are exactly $d^{(k)}-1$ amplitude boundary condition equations at each internal vertex of degree k.  Now sum over all internal vertices.  The sum over k of $d^{(k)}$ is equal to the total number of internal edges that are connected by these vertices, and is equal to $2E-V_{E}$.  The sum of unity over all internal vertices is $V_{I}$.  Thus, there are exactly $2E-V_{E}-V_{I}=2E-V$ total internal amplitude boundary condition equations.  This yields $V+2E-V=2E$ total boundary condition equations, exactly the same number as the number of unknown edge amplitudes.  A nontrivial solution requires the determinant of the coefficients to vanish, leading to the secular equation determining the eigenvalues of the graph, as required.
%%%%%%%%%%%%%%%%%%%%%%%%%%%%%

The transition moments for the graph are calculated using the full eigenstates and are sums over edges of the moments of the edge coordinate times an angular factor describing the geometric position of the edge relative to an external axis used to define the vertices of the graph:

\begin{equation}\label{xNM}
x_{nm}=\sum_{i=1}^{E}\int_{0}^{a_{i}}\phi_{n}^{*i}(s_{i})\phi_{m}^{i}(s_{i})\ x(s_{i})ds_{i} ,
\end{equation}
where $\phi_{m}^{i}(s_{i})$ are the normalized edge wave functions given in Eq. (\ref{edgeFunctions2}). Here, $x(s_{i})$ is the x-component of $s_{i}$, measured from the origin of the graph (and not of the edge), and is a function of the prior edge lengths and angles.  With edge wave functions of the form of Eq. (\ref{edgeFunctions2}), the computation of the transition moments requires integrals of products of sines and cosines with either $s$ or $1$, all of which are calculable in closed form.  Detailed examples for loops, wires, and stars are available in the literature from our prior work\cite{lytel12.01,shafe12.01,lytel13.01}.

For a reference frame that is rotated $\theta$ degrees with respect to the initial reference frame, the diagonal components, $\beta_{xxx}(\theta)$ and $\gamma_{xxxx}(\theta)$, can be determined using
\begin{eqnarray}\label{BetaCartesian}
\beta_{xxx}(\theta) &=& \beta_{xxx}\cos^3\theta \nonumber + 3\beta_{xxy}\cos^2\theta \sin\theta \nonumber \\ &+&  3\beta_{xyy}\cos\theta \sin^2\theta + \beta_{yyy}\sin^3\theta,
\end{eqnarray}
and
\begin{eqnarray}\label{GammaCartesian}
\gamma_{xxxx}(\theta) &=& \gamma_{xxxx}\cos^4\theta+4\gamma_{xxxy}\cos^3\theta\sin\theta + 6\gamma_{xxyy}\cos^2\theta \sin^2\theta\nonumber \\
&+& 4\gamma_{xyyy}\cos\theta \sin^3\theta + \gamma_{yyyy}\sin^4\theta
\end{eqnarray}
where the value of $\theta$ that maximizes the left-hand side of either equation is usually different for each of $\beta_{xxx}$ and $\gamma_{xxxx}$, and the tensor components on the right-hand side of either equation are referenced to the graph's intrinsic frame where the hyperpolarizabilities are calculated.  By definition, $\beta_{xxx}$ ($\gamma_{xxxx}$) is at an extreme value when the graph is rotated through $\theta$.  Once the graph is solved and the tensor components are known in its frame, $\theta$ is easily found by maximizing (\ref{BetaCartesian}) for $\beta_{xxx}$ and (\ref{GammaCartesian}) for $\gamma_{xxxx}$.  The tensor norms are invariant under any transformation and provide immediate insight into the limiting responses of the graphs.  They are given by
\begin{equation}\label{BetaNorm}
|\beta| = \left( \beta_{xxx}^2+3\beta_{xxy}^2+ 3\beta_{xyy}^2+\beta_{yyy}^2\right)^{1/2}
\end{equation}
and
\begin{equation}\label{GammaNorm}
|\gamma| = \left( \gamma_{xxxx}^2+4\gamma_{xxxy}^2+ 6\gamma_{xxyy}^2+4\gamma_{xyyy}+\gamma_{yyyy}^2\right)^{1/2}
\end{equation}
These are the magnitudes of the graph's hyperpolarizabilities and are both scale and orientation-independent.

The use of tensors to extract the nonlinear optical response as a function of geometry and topology is most easily achieved by transforming the reducible Cartesian tensor representations to a set of irreducible spherical tensors.  Molecular nonlinearities in multipolar media have been analyzed using irreducible, spherical representations for $\beta$, an approach which provides insight into the shape-dependence of the first hyperpolarizability, particularly with respect to certain symmetry groups\cite{zyss94.01,joffr92.01}.

For fully symmetric Cartesian tensors, $\beta$ has a vector $(J=1)$ and a septor $(J=3)$ component\cite{jerph78.01}.  Its irreducible representation is $1 \bigoplus 3$, with a total of four independent Cartesian components, as noted earlier.  Similarly, $\gamma$ has a scalar, deviator, and nonor component, and its irreducible representation is $0 \bigoplus 2 \bigoplus 4$.  The specific form of the spherical tensor expansions may be directly calculated using the Clebsch-Gordon coefficients.  Using well established methods\cite{jerph78.01}, we get
\begin{eqnarray}\label{BetaTensors}
S_{\pm1}^{1} &=& \sqrt{(3/10)}\left[\pm \left(\beta_{xxx}+\beta_{xyy}\right) +\imath \left(\beta_{yyy}+\beta_{xxy} \right)\right] \nonumber \\
S_{\pm1}^{3} &=& \sqrt{(3/40)}\left[\pm \left(\beta_{xxx}+\beta_{xyy}\right) +\imath \left(\beta_{yyy}+\beta_{xxy}\right)\right] \\
S_{\pm3}^{3} &=& \sqrt{(1/8)}\left[\pm(-\beta_{xxx}+3\beta_{xyy}) +\imath(\beta_{yyy}-3\beta_{xxy})\right].\nonumber
\end{eqnarray}
Similarly, the spherical components of $\gamma$ are given by
\begin{eqnarray}\label{GammaTensors}
T_{0}^{0} &=& \sqrt{(1/5)}\left[\gamma_{xxxx}+2\gamma_{xxyy}+\gamma_{yyyy}\right] \nonumber \\
T_{0}^{2} &=& \sqrt{(1/7)}\left[\gamma_{xxxx}+2\gamma_{xxyy}+\gamma_{yyyy}\right] \nonumber \\
T_{\pm2}^{2} &=& \sqrt{(3/14)}\left[(-\gamma_{xxxx}+\gamma_{yyyy}) \mp2\imath(\gamma_{xxxy}+\gamma_{xyyy})\right] \nonumber \\
T_{0}^{4} &=& \sqrt{(9/280)}\left[\gamma_{xxxx}+2\gamma_{xxyy}+\gamma_{yyyy}\right] \\
T_{\pm2}^{4} &=& \sqrt{(1/28)}\left[(-\gamma_{xxxx}+\gamma_{yyyy}) \mp2\imath(\gamma_{xxxy}+\gamma_{xyyy})\right] \nonumber \\
T_{\pm4}^{4} &=& \sqrt{1/4}\left[(\gamma_{xxxx}-6\gamma_{xxyy}+\gamma_{yyyy}) \pm 4 \imath (\gamma_{xxxy}-\gamma_{xyyy})\right]\nonumber
\end{eqnarray}
where $\imath^2 = -1$.

The total tensor norm for $\beta$ is found by summing $|S_{m}^{J}|^2$ over the $2m+1$ components $m=-J,-J+1,...0...J-1,J$ for $J=1$ and $J=3$.  Similarly, the total tensor norm for $\gamma$ is found by summing $|T_{m}^{J}|^2$ over the $2m+1$ components $m=-J,-J+1,...0...J-1,J$ for $J=0$, $J=2$, and $J=4$.  These norms are of course identical to those that would be computed from the original Cartesian tensors. But the new information here is that we now have a geometric description of the rotation properties of graphs as a function of their shape that can display their most significant contributions in terms familiar to designers of nonlinear optical molecules.  We return to the interpretation of the hyperpolarizabilities in terms of their spherical tensor representations later in Sec \ref{sec:delta3Atom}.  With this machinery, we are able to begin the analysis of the dressed quantum graphs.

\section{The delta wire graph motif}\label{sec:deltaWires}

Dressed quantum graphs are defined by the existence of a self-adjoint Hamiltonian with a nonzero potential energy.  For delta function potentials, the self-adjoint requirement creates a boundary condition at the location of the delta function $V(s)=V_{0}\delta(s-a)$ that is a generalized version of the canonical condition $\phi^{'}(a+\epsilon)-\phi^{'}(a-\epsilon)=2V_{0}\phi(a)$ as $\epsilon\rightarrow 0$ for the discontinuity in the derivative of the edge function on either side of the delta function.  The wavefunction is continuous across the location of the delta function.  The generalized version for $N$ edges emanating from a single vertex where a delta function potential $(g/L)\delta(s-a)$ is placed, each supporting a set of edge functions $\phi_{n}^{i}$, takes the form
\begin{equation}\label{genBC}
\sum_{i=1}^{N}\phi_{n}^{'i}=2(g/L)\phi_{n}(a) ,
\end{equation}
where the derivative on each segment, $\phi_{n}^{'i}$, is evaluated along a path starting at the vertex.  The function on the right $\phi_{n}(a)=\phi_{n}^{i}(a)$ is the identical value of each edge's wavefunction at the vertex.  We have chosen units such that $\hbar=m=1$ so that energies have dimensions of $1/length^2$.  This makes the vertex strength g dimensionless.  Further, we have normalized the delta function to a length L which is defined by the requirement that L scales identically with the scaling factor of any other length in the graph.  For example, for $N$ edges, each of length $L_{i}$, we can define $L\equiv\sum_{i=1}^{N}L_{i}$.  This makes the problem scale-invariant.

We may solve any graph with a delta potential by matching the boundary conditions and demanding a unique solution.  As previously discussed\cite{lytel13.01}, this results in a secular equation for the wavenumbers $k_{n}$ which is generally transcendental and will depend on the length scale and the dimensionless delta function strength, g.

Though it is straightforward to solve the quantum wire with a single delta function by matching the discontinuity, it is useful and illustrative for our purposes to invoke the motif method for computing secular equations of complex graphs\cite{lytel13.02}.  This method is a set of rules for writing down by inspection the secular equation for a complex graph using the fundamental expressions for the secular functions controlling the flux in elemental (motif) graphs, such as the star or lollipop graphs.  We employ it here to solve the wire graph with a finite delta function potential and extend it to the two- and three-delta function potential graphs.  Generalization to N delta functions is straightforward, albeit a bit messier.

\begin{figure}\center
\includegraphics[width=3.4in]{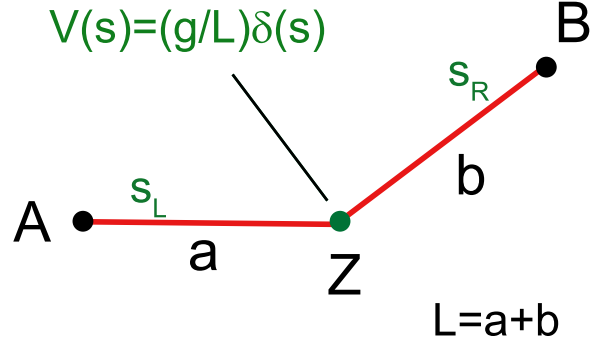}
\caption{Motif graph for a two wire with a delta potential at their intersection vertex.}\label{fig:deltaMotif}
\end{figure}

The motif for a bent delta wire is shown in Fig \ref{fig:deltaMotif}.  This type of graph is often called the compressed, one-dimensional delta atom\cite{blume06.01}.  The potential may be written as
\begin{equation}\label{deltaPot}
V(s)=(g/L)\delta(s)
\end{equation}
where a is the length of the left edge, b is the length of the right edge, and $L=a+b$ is the total length of the graph.  Here, s is the distance measured from the location of the delta function on each edge.  The dimensionless strength parameter g is positive for a potential barrier and negative for a well at the vertex Z.  We demand that the edge functions match at $s=0$, have fluxes which satisfy Eq. (\ref{genBC}) at $s=0$, and have values equal to A and B at the left and right edges, respectively.  These endpoints are able to transfer flux to edges connected to them.  If they are unconnected, $A=B=0$, and the edge functions vanish at the endpoints (corresponding to an infinite potential barrier).  Since each edge points outward from the vertex Z, the edge functions may be written as
\begin{eqnarray}\label{edgeFunctions}
\phi_{L}(s) &=& \frac{Z\sin{k(a-s_{L})}+A\sin{ks_{L}}}{\sin{ka}} \\
\phi_{R}(s) &=& \frac{Z\sin{k(b-s_{R})}+B\sin{ks_{R}}}{\sin{kb}}\nonumber
\end{eqnarray}
which is a canonical form that automatically matches edge functions at the vertex.  Note that if either denominator in Eq.
(\ref{edgeFunctions}) is zero, continuity demands that both are zero, and in this case, the slope is continuous, the effect of the delta function is absent, and the graph becomes a straight wire.  These cases are simple to deal with in a numerical simulation.

Applying the boundary condition Eq. (\ref{genBC}), it is straightforward to show that the relationship among the amplitudes is
\begin{equation}\label{secReln}
ZF_{\delta}=A\sin{kb}+B\sin{ka}
\end{equation}
where the secular function for this graph is
\begin{equation}\label{secPos}
F_{\delta}\equiv F_{\delta}(g;\omega,kL) = -(1/kL)\left[g(\cos{kL}-\cos{\omega kL})-kL\sin{kL}\right]
\end{equation}
Note that we have shifted from using the two edge lengths $a,b$ to the total length L and to $\omega=2a/L-1$, a dimensionless parameter ranging from -1 to 1 describing the asymmetry of the location of the potential along the wire \cite{blume06.01}.  Eq. (\ref{secReln}) expresses the flow of flux in the graph when the ends are sources or sinks of flux from other edges to which they are connected.  This is the motif for the two wire graph with a single, finite delta function potential at its vertex.  We will next invoke the motif to solve the compressed 1-delta, 2-delta, and 3-delta wire graphs.

\section{The compressed delta atom}\label{sec:deltaAtom}

The delta graph motif described above becomes a real quantum graph with vanishing wavefunctions at the ends when $A=0$ and $B=0$.  Then the solutions to $F_{\delta}=0$ yield the eigenvalues $k_n$ of the graph.  The energies are then $E_n=k_n^2/2$ in units where $\hbar=m=1$.

The positive energy solutions $k_nL$, when the wavefunctions are not localized on the delta function, are real numbers whose values fall between root separators located at $n\pi$.  This is true regardless of the sign of g.  For finite g, the eigenstates are nondegenerate.  The negative energy solutions associated with localization of the wavefunction at the delta function may be sought by replacing $kL$ with $i\kappa L$, where $\kappa$ is real. This changes Eq. (\ref{secPos}) to
\begin{equation}\label{secNeg}
F_{\delta}(g;\omega,\kappa L) = (i/\kappa L)\left[g(\cosh{\kappa L}-\cosh{\omega\kappa L})+\kappa L\sinh{\kappa L}\right]
\end{equation}
The first term on the right hand side of Eq. (\ref{secNeg}) always has the sign of g, since the factor multiplying it is always positive.  Since the second term is always positive, the potential with $g>0$ has no negative energy solutions.  For $g<0$, there is exactly one negative energy solution but only if $g<g_c$, where $g_c=2/(\omega^2-1)$ is the (negative) threshold above which no negative energy state may exist.  This may be seen by expanding $F_{\delta}$ as a power series in $x\equiv\kappa L$ which yields
\begin{eqnarray}\label{secNegExp}
-\imath xF_{\delta}(g;\omega,\kappa L) &=& x^{2}\left[1-g\left(\frac{\omega^{2}-1}{2}\right)\right]+O(x^{4})\nonumber \\
&=& x^{2}\left[1-\frac{g}{g_c}\right]+O(x^{4})
\end{eqnarray}
where the definition of $g_c$ has been used.  This is the leading term when $x\ll 1$ and is negative for $g<g_{c}<0$.  But as $x$ increases, the right hand side always turns positive.  Hence, there is a single, negative root whenever $g<g_{c}$.

As negative $g$ gets large, the value of $x\equiv\kappa L$ that satisfies Eq. (\ref{secNeg}) gets very large and approaches $|g|$, since the hyperbolic functions both scale the same way for large arguments.  This implies that the ground state energy gets larger and more negative as the well gets deeper.

\begin{figure}\center
\includegraphics[width=5in]{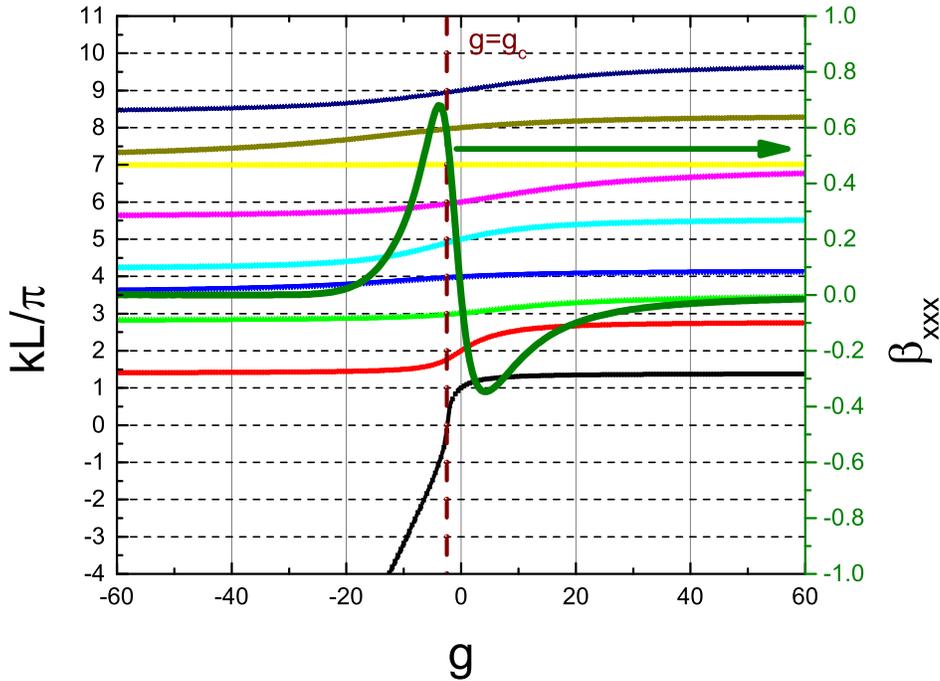}
\caption{Composite graph showing the variation of the eigenspectrum (left axis) as g varies from large negative to large positive values.  The $\delta$ function is located at $\omega=-.44\ (a=.28)$, for which $g_{c}=-2.48$. The root separators are shown as dashed lines.  The negative energy state appears for $g<g_{c}$.  The variation of $\beta_{xxx}$ with g is shown on the right axis.  The composite graph shows that the largest $\beta_{xxx}$ occurs for negative g, suggesting that the appearance of a negative energy state is responsible for the large hyperpolarizability $\beta_{xxx}=0.68$.}\label{fig:1delta_spectra_beta_VS_g}
\end{figure}

\begin{figure}\center
\includegraphics[width=5in]{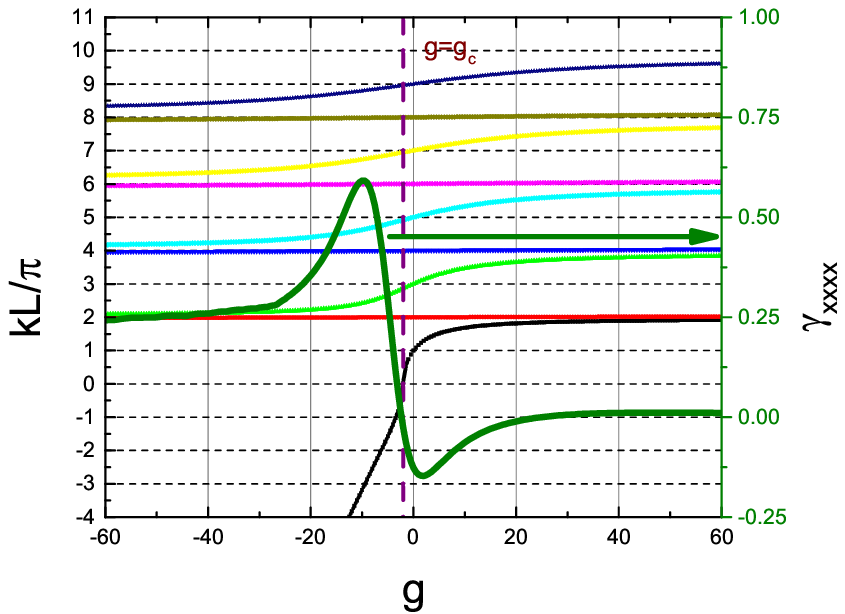}
\caption{Composite graph showing the variation of the eigenspectrum (left axis) as g varies from large negative to large positive values.  The $\delta$ function is located at $\omega=.02\ (a=.51)$, for which $g_{c}=-2.00$. The root separators are shown as dashed lines The negative energy state appears for $g<g_{c}$.  The variation of $\gamma_{xxxx}$ with g is shown on the right axis.  The composite graph shows that the largest $\gamma_{xxx}$ occurs for negative g.}\label{fig:1delta_spectra_gamma_VS_g}
\end{figure}

Figures \ref{fig:1delta_spectra_beta_VS_g} and \ref{fig:1delta_spectra_gamma_VS_g} show the first nine solutions as a function of g, along with $\beta$ and $\gamma$, respectively.  The roots $k_{n}$ are separated by fixed root boundaries, and all shift up one boundary when the negative energy state disappears (becoming the positive ground state).  It is notable that $\beta$ and $\gamma$ both achieve their large values when the well gets deep and the ground state energy goes negative large.

Once the vertices and delta potential strength g are specified, the eigenvalues may be computed from Eq. (\ref{secPos}) and Eq. (\ref{secNeg}), and the edge functions in Eq. (\ref{edgeFunctions}) become fully determined, once the overall wavefunction is normalized.  The transition moments may then be computed from Eq. (\ref{xNM}), and the hyperpolarizability tensors may be computed using Eq. (\ref{BetaCartesian}) and Eq. (\ref{GammaCartesian}).  Their values and ranges depend upon the strength g of the potential and its position on the straight wire.

\begin{figure}\center
\includegraphics[width=5in]{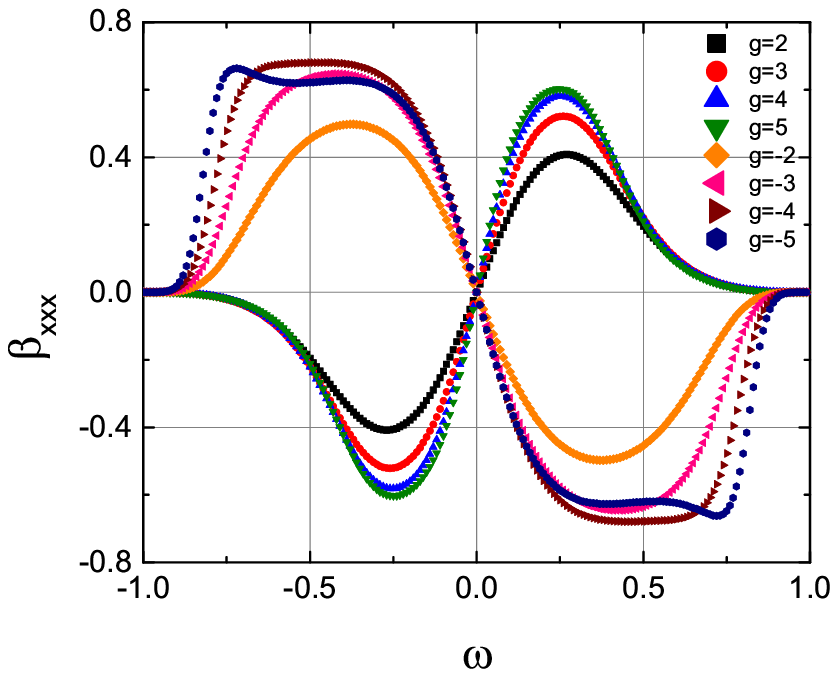}
\caption{$\beta_{xxx}$ as a function of the position of the delta function for sets of graphs with positive and negative g.}\label{fig:1delta_linePlotsVSg_beta}
\end{figure}

Figure \ref{fig:1delta_linePlotsVSg_beta} displays the variation of $\beta_{xxx}$ as the location of the delta function moves from one end to the other for several values of both positive and negative strength g.  It is seen that $\beta_{xxx}$ changes sign as the delta function moves through the center and the graph remains fixed.  Positive and negative g profiles are of opposite sign.  For either sign, $\beta_{xxx}$ increases with g up to a point, then decreases.  For $g>0$, the optimum is $g \approx 3.5$.  For negative g, the scaling behavior is more complex.  The maximum value of $\beta_{xxx}$ remains constant at about $0.68$ but position of the delta function that yields it is pushed toward the edge.  In fact, the $g=-5$ curve seems to rise to its maximum as the delta moves from the left edge, but then it drops to a lower value, rather than increasing and perhaps exceeding the $0.71$ limit.  This effect appears to be related to the universal result that all graphs whose $\beta$ are calculated by starting with a Hamiltonian and a potential of any kind have $\beta_{xxx} \leq 0.7089$, the potential optimization limit\cite{zhou07.02,ather12.01}.  The sum rules do not constrain $\beta_{xxx}$ from equaling unity, as has been shown in a Monte Carlo calculation that starts with the energies and transition moments (top-down), rather than with a Hamiltonian and its energies and states (bottom-up)\cite{kuzyk09.01}.  But the curve is a compelling reminder that there is some physical constraint at work for conservative, self-adjoint Hamiltonian potential systems that limits the intrinsic response to about $0.7089$, rather than unity.  As discussed elsewhere, there may exist non-standard Hamiltonians that span the gap between the potential optimization limit and unity, but none have been discovered to date\cite{watki12.01}.  It is gratifying that the quantum wire with a single, finite delta potential well exhibits a nonlinearity equaling the potential optimization limit.

\begin{figure}\center
\includegraphics[width=5in]{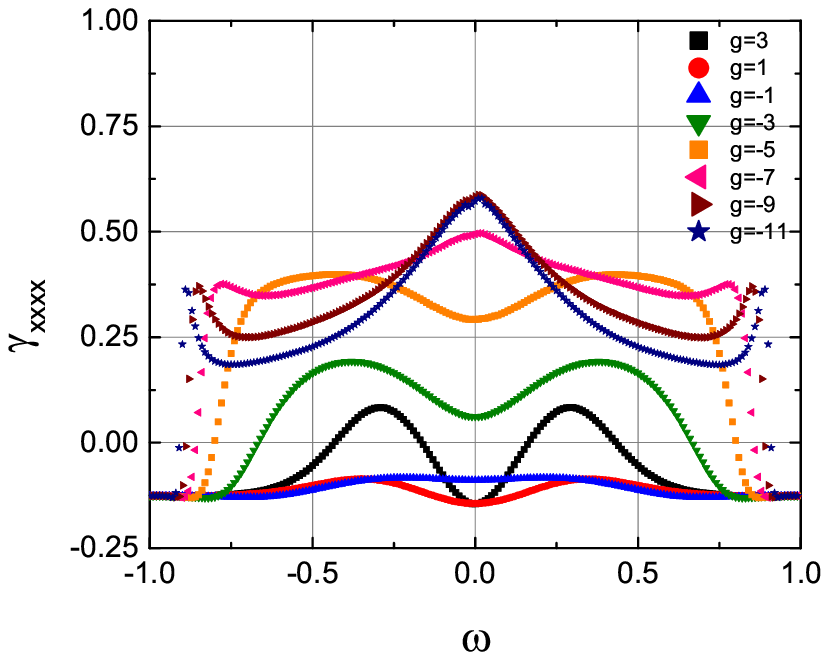}
\caption{$\gamma_{xxxx}$ as a function of the position of the delta function for sets of graphs with positive and negative g.}\label{fig:1delta_linePlotsVSg_gamma}
\end{figure}

Figure \ref{fig:1delta_linePlotsVSg_gamma} displays the variation of $\gamma_{xxxx}$ as the delta function is moved from one end of the wire to the other.  For $0<g<2$, $\gamma_{xxxx}$ is negative over the wire, but positive values appear for larger g at specific locations on the wire.  For negative g, $\gamma_{xxxx}$ is positive over most of the well and increases with increasing g, up to sixty percent of the theoretical limit of unity.  The lower bound on the graphs is also sixty percent of the theoretical lower limit of $-0.25$.

\begin{figure}\center
\includegraphics[width=5in]{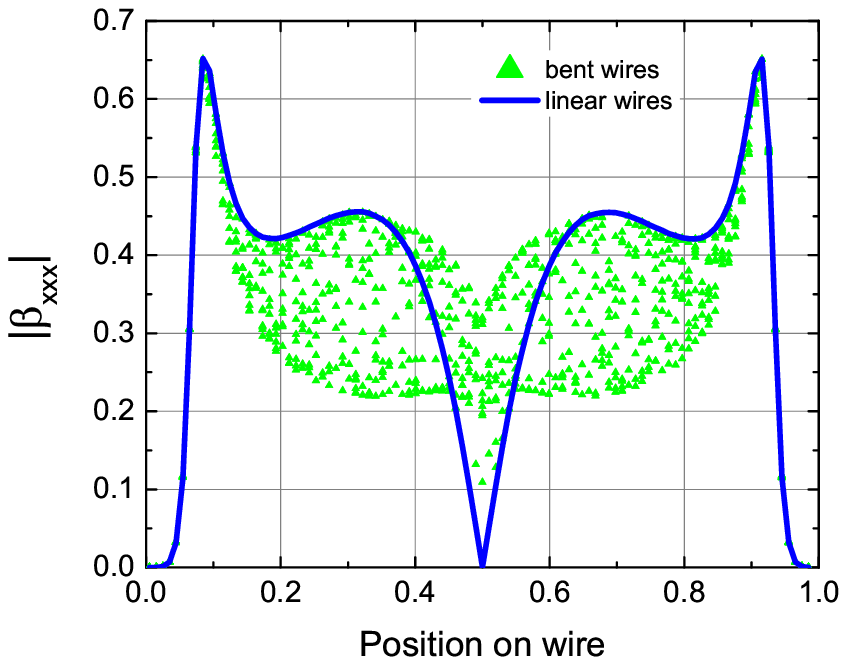}
\caption{$\beta_{xxx}$ as a function of the position of the delta function on the wire for $g=-7$ for linear 2-wires (blue) and bent 2-wires (green).  The bent wires are an ensemble of graphs with randomly chosen angles between the edges, but with the graph rotated to yield the maximum value of $\beta_{xxx}$ in the lab frame.  For clarity, we plot the absolute value of $\beta_{xxx}$ to make clear the relation between the two plots.}\label{fig:betaDeltaAngAndLinear}
\end{figure}

Figure \ref{fig:betaDeltaAngAndLinear} compares the values of $\beta_{xxx}$ as the delta function moves across the wire for straight wires and for wires with two bent segments connected at the delta vertex.  The figure has been displayed in a symmetric way for clarity, but the values to the right of center are actually the negative mirror of the ones to the left if the wire remains unrotated as the $\delta$ function moves right past the center of the wire.  The Monte Carlo calculation that generated these results selected the relative angle at random, calculated the hyperpolarizability tensor components, then used the tensor properties\cite{lytel13.01} of $\beta_{ijk}$ to rotate the wire into its preferred position (where $\beta_{xxx}$ is maximum in the lab frame).

Next, we turn to the scaling of the hyperpolarizabilities and the three-state model parameters X and E as a function of the topological properties of the compressed delta atom graphs.  We start by investigating how many significant levels contribute to $\beta$ and $\gamma$ as the parameters g and $\omega$ are varied from values where the hyperpolarizabilities are small to values near the fundamental limits.  An empirical rule of thumb regarding the fundamental limits of nonlinear optical structures is that the contributions to $\beta$ arise from three states only as $\beta$ approaches the limit\cite{kuzyk00.01}.  This rule is satisfied for every structure examined to date, and though a fundamental proof remains elusive, a recent analysis asserts that it is likely that this three-level Ansatz (TLA) is always true for $\beta$, implying that a three level model suffices to describe the fundamental limit of the first hyperpolarizability for any material\cite{shafe13.01}.  Prior work has shown that the TLA holds for the second hyperpolarizability, as well, though there has been speculation that for certain structures, a four-level model of the fundamental limit may be necessary.\cite{watki12.01}

To investigate, we create an ensemble of graphs where $\omega$ varies from -1 to 1 in 100 steps for each of 50 values of g varying from -12 to +12.5 in steps of 0.5.  For each of the 5000 graphs, the hyperpolarizabilities were calculated, and their densities were also calculated in order to determine the ten most important states for each.  The truncated hyperpolarizabilities $\beta_{3}$ and $\beta_{4}$ ($\gamma_{3}$ and $\gamma_{4}$) were then calculated for $\beta$ ($\gamma$) using only the three and four most important states, respectively.  The truncated hyperpolarizabilities are then compared with the full values.

\begin{figure}\center
\includegraphics[width=5in]{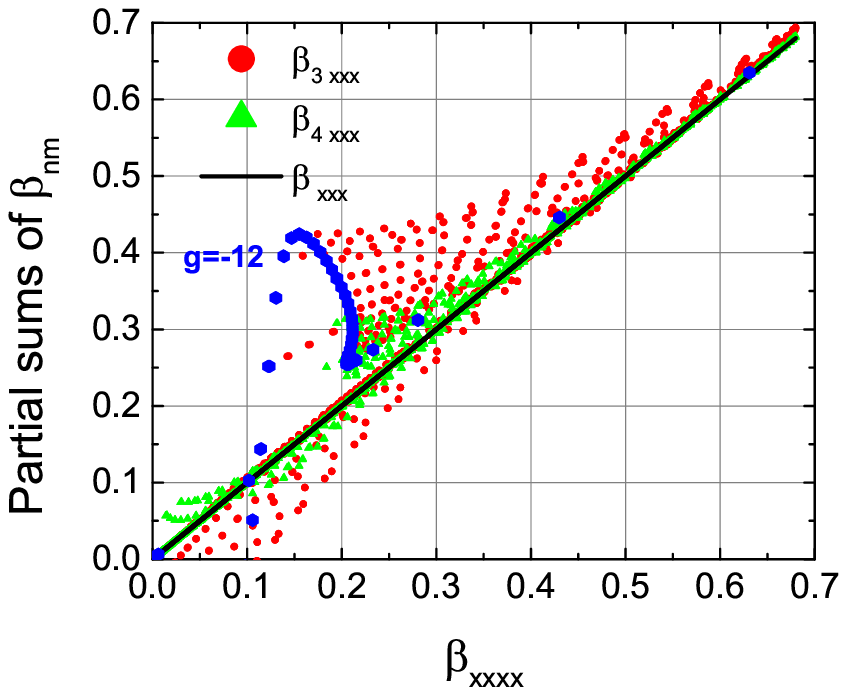}
\caption{Scatterplot of partial sums of $\beta_{xxx}$ for the three and four most significant states for a dataset that includes 100 points across the wire and 50 values of the strength g. The contributions from the graphs having $g=-12$ are highlighted with large blue hexagons.}\label{fig:1delta_beta_TLA}
\end{figure}

Fig \ref{fig:1delta_beta_TLA} displays a scatterplot of $\beta_{3}$, $\beta_{4}$, and $\beta$ for the 5000 graphs. Each circular point corresponds to the 3-state value, i.e. where only the three most significant states are included in the partial sum, while the triangular points are a partial sum of 4 states.  The rising, curved sets of circles each correspond to a unique value of $g$ but varying position of the delta function; those points with $g=-12$ are called out using pentagons.  A single, solid line labeled $\beta_{xxx}$ is the result for all 5000 graphs when all states are included.  As $g$ increases, the rising, curved sets of circles move to the upper right, with their larger values getting closer to the exact values.  At the value of $g\approx -4$, the three-level approximation is nearly equal to the four-level approximation, and both are nearly identical to the exact value of $\beta$.  This is the manifestation of the three-level Ansatz for $\beta$.  Note, too, that the four-level sums are always close to the exact values, though there is some fluctuation until the maximum value of $\beta$ is achieved.

\begin{figure}\center
\includegraphics[width=2.5in]{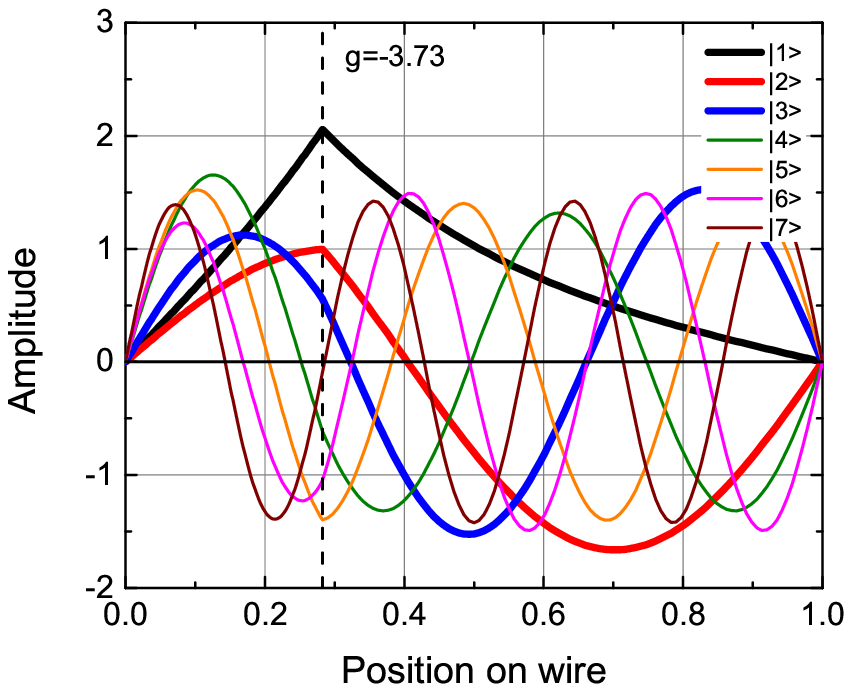}\includegraphics[width=2.5in]{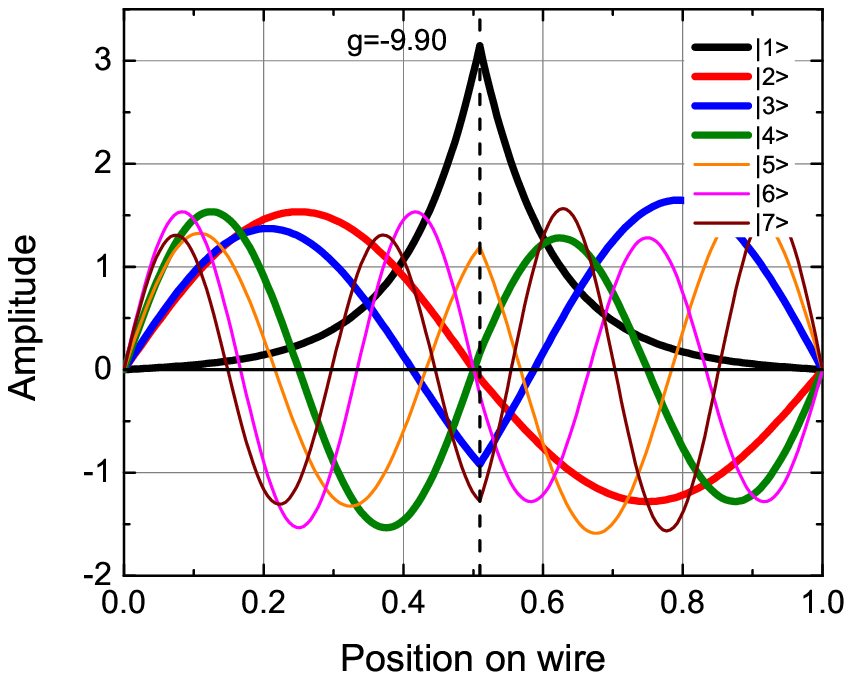}\\
\includegraphics[width=2.5in]{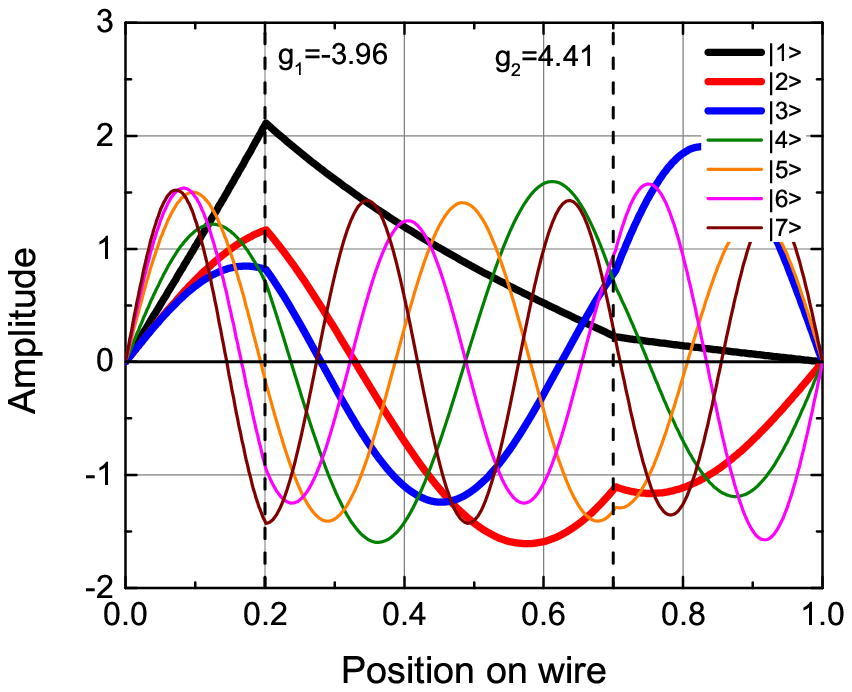}\includegraphics[width=2.5in]{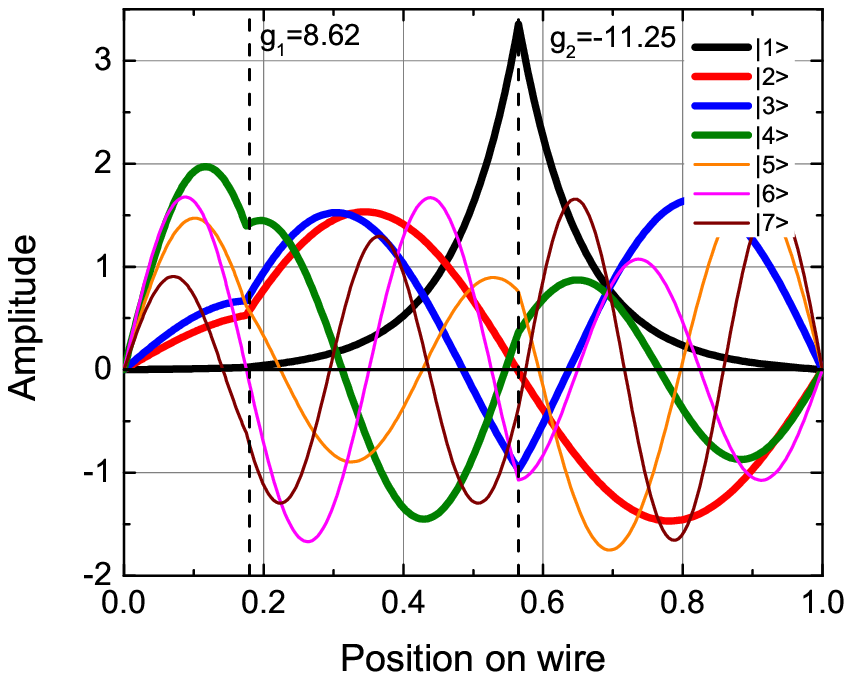}\\
\includegraphics[width=2.5in]{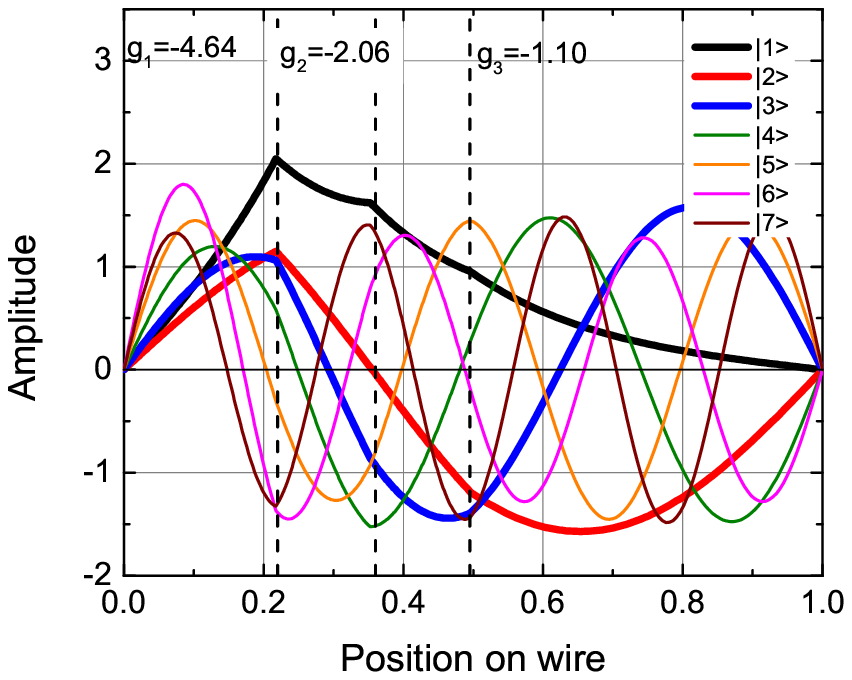}\includegraphics[width=2.5in]{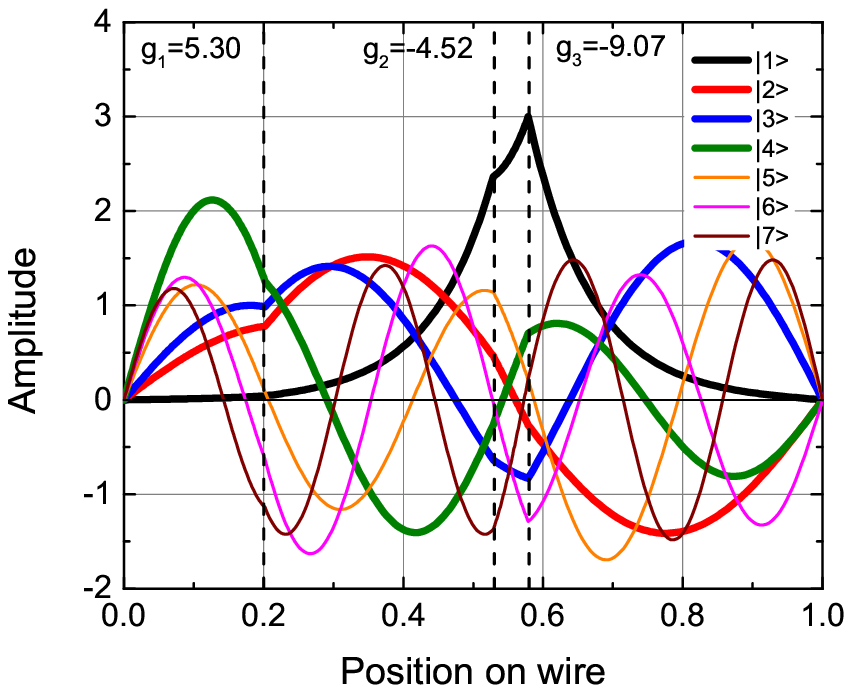}\\
\caption{First seven eigenstates of the compressed $\delta$ atom and molecule graphs with maximum $\beta_{xxx}$, with one, two, and three $\delta$ functions (left).  The maxima for $\beta$ are $0.680, 0.697$, and $0.709$ respectively.  The figures on the right are the eigenstates with maximum $\gamma_{xxxx}$ with one, two, and three $\delta$ functions.  The maxima for $\gamma_{xxxx}$ are $0.058$, $0.588$, and $0.59$ respectively.}\label{fig:123delta_states_bestBetaGamma}
\end{figure}

The exact value of g for which the maximum of $\beta\approx 0.68$ is achieved is $g=-3.73$.  The eigenstates for this maximum graph are displayed in the top left panel of Fig \ref{fig:123delta_states_bestBetaGamma}, where the three most important states are highlighted with thicker lines than the next four most important states.  For later use, we have placed this figure in a collage with similar figures for the two-delta and three-delta graphs.

The optimum position of the $\delta$ function is evident, as well. The first three states have the most significant overlap compared to that of any of them with the remaining states.  Furthermore, the wavefunctions of the higher-energy states, though overlapping with the lower-energy states in the wings, give small transition moments to the lower-energy states by virtue of the oscillations.  In fact, it is seen that the sum over states of $\beta_{nm}$ with $n,m>3$ is likely to have many cancelations, resulting in a near-zero value, as predicted by the TLA.  There is no formal proof of the TLA, but it seems there is a very close link between its validity and the potential optimization maximum of $0.7089$.

The impact of the $\delta$ potential on the value of $\beta$ is profound, because when $g=0$, the potential is absent, and the value of $\beta$ is exactly zero.  The potential breaks the symmetry of the compressed delta atom graph and enables $\beta$ to take a very large value by allowing the first three states to have discontinuous slopes, thus maximizing their overlap.  A quantum wire with a point defect representing a potential well should show similar enhancement of its nonlinearity.  This observation suggests that experimental efforts to construct wires with point defects could lead to highly active materials, though proof requires both an experimental demonstration and a more realistic quantum model with several electrons or bands.

\begin{figure}\center
\includegraphics[width=5in]{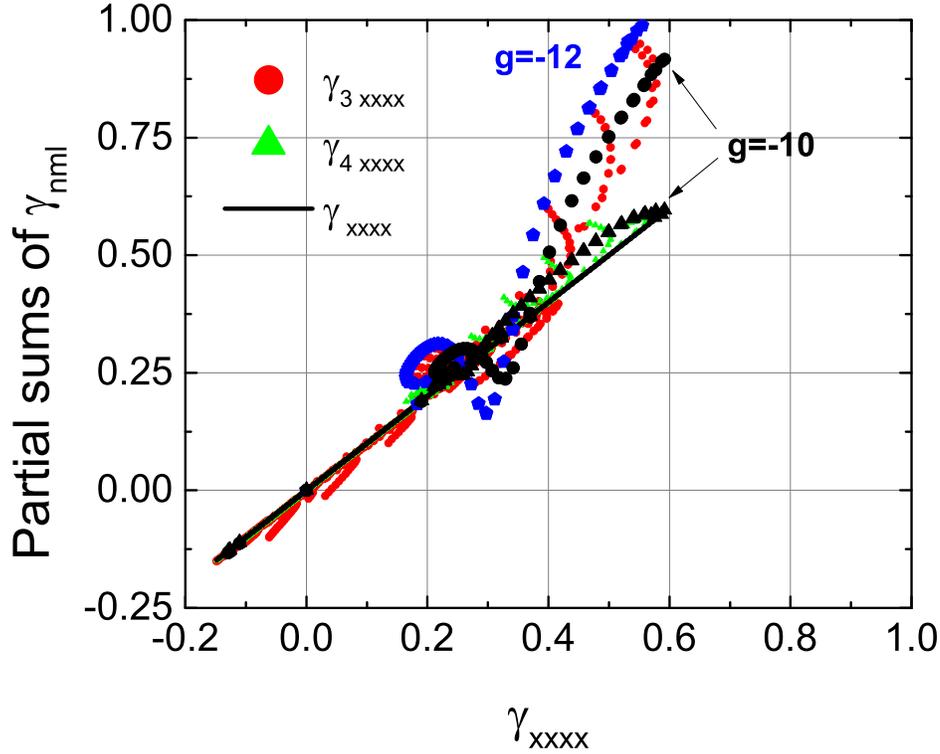}
\caption{Scatterplot of partial sums of $\gamma_{xxxx}$ for the three and four most significant states for a dataset that includes 100 points across the wire and 50 values of the strength g.  The contributions from the graphs having $g=-12$ are highlighted with blue pentagons.}\label{fig:1delta_gamma_TLA}
\end{figure}

Fig \ref{fig:1delta_gamma_TLA} displays a scatterplot of $\gamma_{3}$, $\gamma_{4}$, and $\gamma$ for the 5000 graphs. Each circular point again corresponds to the 3-state value, while the triangle points are the 4-state values.  The behavior shown in this figure is quantitatively different than that for $\beta$ show in Fig \ref{fig:1delta_beta_TLA}.  This time, the rising sets of circles that bend left correspond to different values of g for the same locations on the wire.  Values shown by pentagons correspond to $g=-12$ and are highlighted for both the three level sum and asymptote to unity.

For $\gamma$, the three level Ansatz never holds, and a fourth state is essential to make the sum converge toward the actual values (solid line).  Even then, convergence only occurs near the maximum value of $\gamma$.  The squares are the values of $\gamma_{3}$ and $\gamma_{4}$ for $g=-10$, the value for which $\gamma$ assumes its maximum value and shows that the four level hypothesis holds exactly for $\gamma$ in this limit.  We conclude that four levels are needed to describe $\gamma$ as it approaches the local maximum values observed in potential optimization studies.\cite{watki12.01}  And as $\gamma$ decreases from its maximum value, four states are not enough for an accurate measure of $\gamma$ unless it is near-zero.  In contrast, three states are sufficient for desribing negative gamma.  It must be noted that the three-level ansatz, which was used to calculate the limits, gives an upper bound that is never found to be breached.  As such, the need for 4 states to describe the second hyperpolarizability near the local maximum of $\gamma$ does not contradcit the fact that three states are needed to calculate the limits at the global maximum.

The exact value of g for which the maximum of $\gamma\approx 0.58$ is achieved is $g=-9.90$.  The eigenstates for this maximum graph are displayed in the top right panel of Fig \ref{fig:123delta_states_bestBetaGamma}, where the four most important states are highlighted with thicker lines than the next three most important states.  The optimum position of the $\delta$ function is evident, as well. The first four states have the most significant overlap compared to that of any of them with the remaining states.  However, the function of the fourth most important state is evidently to reduce the partial sum over states from its higher three-level value (c.f. Fig \ref{fig:1delta_gamma_TLA}) to a value closer to the exact sum.  This is evident from the asymmetry of the fourth state relative to the first and third states.  For $\gamma$, it is true that three states overestimate the nonlinearity, and a fourth state is always required to get to the exact value at the maximum.  The need for four states arises from the distribution of states in the wire as originating in the quantum mechanics of the eigenstates and the discontinuity offered by the $\delta$ potential.

The three-level model for $\beta$ contains the two parameters $X=x_{01}/x_{01}^{max}$ and $E=E_{10}/E_{20}$, with $x_{01}^{max}$ given by Eq. (\ref{Xmax}), with $r\rightarrow x$, since the graphs are aligned along the x-axis.  When the TLA holds for $\beta$, X approaches $0.79$, while E approaches $0.5$ for all known potential models whose truncated sum rules work for three states.  The three level model for $\beta_{xxx}$ is
\begin{equation}\label{beta3Lmodel}
\beta_{xxx}^{3L} = 3^{3/4}E_{10}^{7/2} \left[\frac{|x_{01}|^{2}\bar{x}_{11}}{E_{10}^{2}} + \frac{|x_{02}|^{2}\bar{x}_{22}}{E_{20}^{2}}+\left(\frac{x_{01}x_{20}x_{12}}{E_{10}E_{20}}+c.c.\right)\right]
\end{equation}
and is valid when the graph has the correct topology and geometry to allow $\beta_{xxx}$ to achieve the maximum value, per potential optimization.  This is the TLA in action.  However, the delta graphs do not exhibit quite the same universal scaling behavior in E and X as all other potentials in optimization studies. In those studies, the sum rules only require three levels to converge, and the use of the three-level sum rules in the three-level expression for $\beta_{xxx}$ enables it to take the form $\beta_{xxx}=f(E)G(X)$, where
\begin{equation}\label{eff}
f(E)= (1-E)^{3/2}\left(E^2+\frac{3}{2}E+1\right)
\end{equation}
and
\begin{equation}\label{Gee}
G(X) = 3^{1/4}X\sqrt{\frac{3}{2}(1-X^4)}
\end{equation}
For all prior potential optimization studies in one dimension, the universal values $E=0.5$ and $X=0.79$ are achieved when $\beta_{xxx}=0.7089$.  The value of $f(.5)G(.79)=0.71$, so the TLA, coupled with the three-level sum rule, is exact in this limit.

\begin{figure}\center
\includegraphics[width=5in]{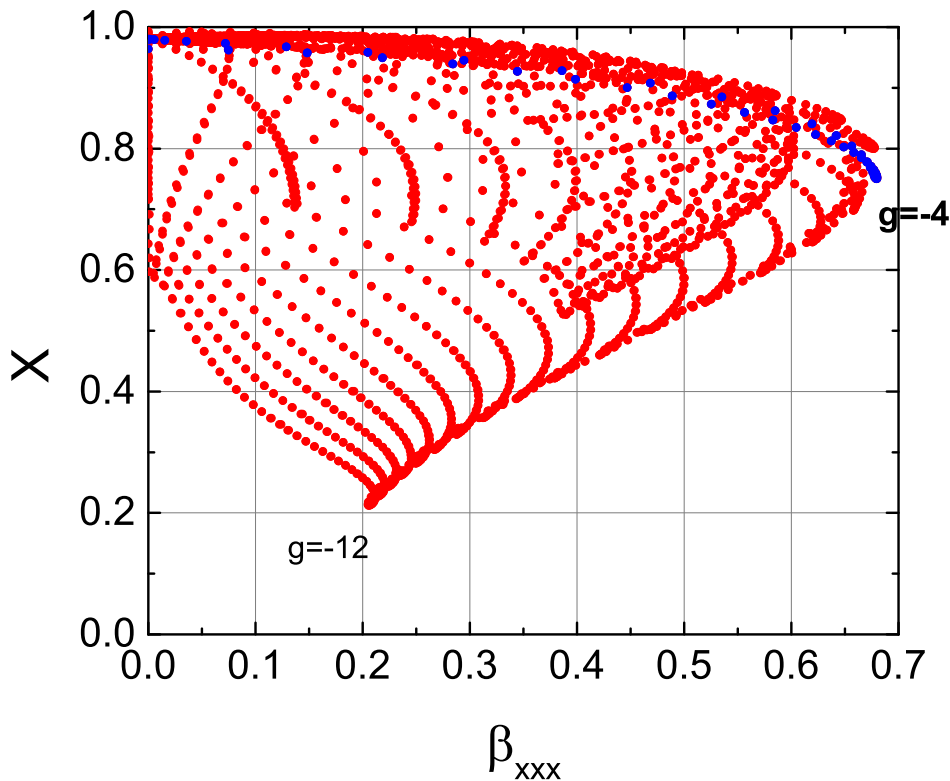}
\caption{Scatterplot of the value of X for the same data set used in the prior two figures.  The contributions from the graphs with $g=-4$ are highlighted in blue.  This is (nearly) the value of g for which the maximum $\beta_{xxx}$ can occur.  Note that X converges to its universal value of about 0.79 when $\beta_{xxx}$ is at its maximum.}\label{fig:1delta_X_vs_beta}
\end{figure}

\begin{figure}\center
\includegraphics[width=5in]{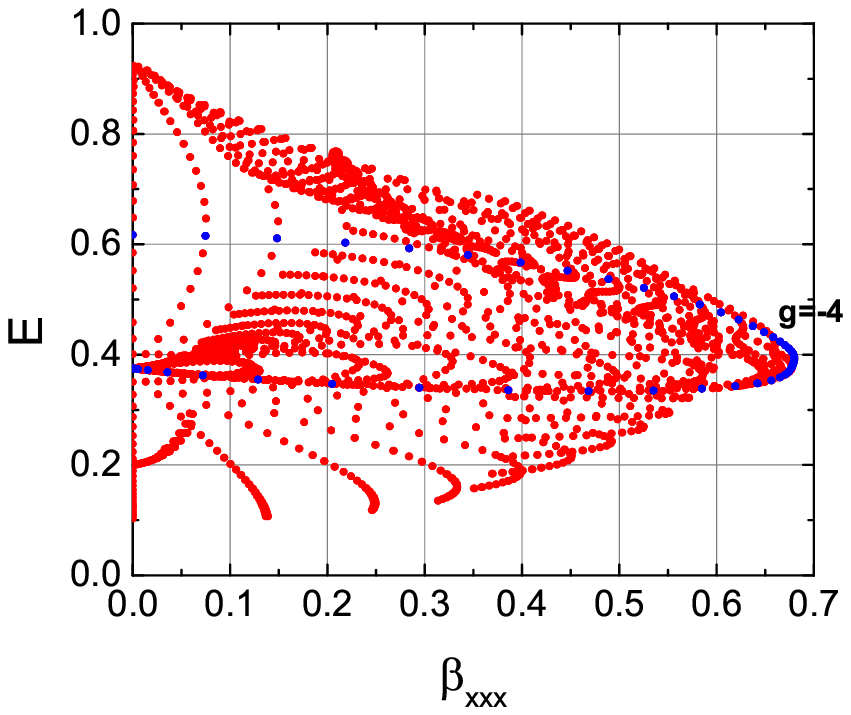}
\caption{Scatterplot of the value of E for the same data set used in the prior three figures.  The contributions from the graphs with $g=-4$ are highlighted in blue.  This is (nearly) the value of g for which the maximum $\beta_{xxx}$ can occur.  Note that E converges to its universal value of about 0.39 when $\beta_{xxx}$ is at its maximum, well below the universal value of $0.5$ observed in prior potential optimization studies.}\label{fig:1delta_E_vs_beta}
\end{figure}

For the delta graphs, we find slightly different scaling behavior for the first time.  This is illustrated for the compressed $\delta$ atom graphs for X and E in Figs \ref{fig:1delta_X_vs_beta} and \ref{fig:1delta_E_vs_beta}, respectively.  The values corresponding to $g=-4$ are highlighted and show that this topology is the one with the maximum value of $\beta$, as predicted by the TLA.  The three level model parameters take on the values $E\sim0.4$ and $X\sim0.79$, yielding a product $f(E)G(X)\sim0.8$, indicating a failure of the three level sum rule for delta graphs.  Careful calculation reveals at least five levels are required to make the sum rule (nearly) converge in this limit.

This result is intriguing precisely because it is an exception to a heretofore universal property established by repeated simulations.  We speculate that it is due to the discontinuous potential energy of the graph, compared to those continuous potentials used in previous studies.

The analysis of the compressed $\delta$ atom graphs has shown for the first time that dressed quantum graphs (dynamical, quasi-1D confined quantum systems) have first and second hyperpolarizabilities that are nearly equal to the fundamental potential optimization limits for specific topologies.  The scaling properties of $\beta$ are reflected by the three-level model scaling parameters, and for $\beta$, the TLA appears to be an exact result.  For $\gamma$, four states are needed to describe the second hyperpolarizability at the limit.  That fact that three states are needed for $\beta$ and four states for $\gamma$ were shown to arise from the exquisite tunability of eigenstate shape and overlap with other eigenstates offered by the finite $\delta$ potential.  The results motivate an experimental search for confined, 1D systems having finite, controllable point defects along their main charge transfer paths.

There remains the question of whether any compressed $\delta$ atom graphs exist that have exactly the maximum values of the hyperpolarizabilities set by the fundamental limits and also the exact, universal value of E at that limit. A very large Monte Carlo run was executed for these graphs with the result that $\beta<0.7$ and $\gamma<0.6$ for all of them, with $E<0.43$.  Since eigenstate overlap tuning was enhanced by a single $\delta$ potential, it is of great interest to investigate what happens when additional finite $\delta$ potentials are incorporated.  These compressed $\delta$ molecules are discussed next, where we show that the maximum hyperpolarizabilities may be exactly achieved, but the value of E remains an exception to the (prior) universal scaling observed in all other potentials studied to date.

%%%%%%%%%%%%%%%%%%%%%%%%%%%%%%%%%%%%%%%%%%%%%%%%%%%%%%%%%%%

\section{Compressed delta atom with two potentials}\label{sec:delta2Atom}

\begin{figure}\center
\includegraphics[width=3.5in]{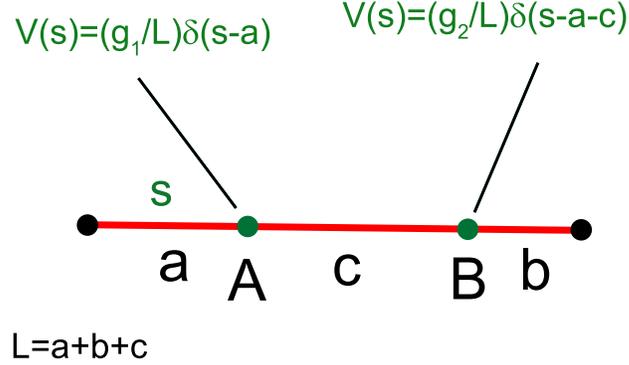}
\caption{Linear wire graph with two delta functions is comprised of two 2-wire delta motifs.}\label{fig:delta2delta}
\end{figure}

Consider now the use of the delta motif to create a graph with two delta potentials on a wire, as in Fig \ref{fig:delta2delta}.  Using the secular expression for the motif in Eq. (\ref{secReln}), we may immediately write down the coupled amplitude equations as
\begin{eqnarray}\label{ampEqns}
A F_{\delta}(g_{1};a,c) &=& B\sin{ka} \\
B F_{\delta}(g_{2};b,c) &=& A\sin{kb}\nonumber
\end{eqnarray}

A subtle but important point is that we need to modify the secular function of the one-delta graph slightly by replacing $g_i/L_i$ with $g_i/L$ ($i=1,2)$ where $L=a+b+c$ is the total, fixed length of the 2-wire system.  This allows the heights of the two delta functions to remain fixed even as they are moved around within the well.  This changes the secular functions to
\begin{eqnarray}\label{secPos1}
F_{\delta}(g_{1};\omega_{1},kL) &=& -(1/kL)\left[g_{1}(\cos{kL_{1}}-\cos{\omega_{1} kL_{1}})\right.\nonumber \\
&-& \left. kL\sin{kL_{1}}\right]
\end{eqnarray}
and
\begin{eqnarray}\label{secPos2}
F_{\delta}(g_{2};\omega_{2},kL) &=& -(1/kL)\left[g_{2}(\cos{kL_{2}}-\cos{\omega_{2} kL_{2}})\right.\nonumber \\
&-& \left. kL\sin{kL_{2}}\right]
\end{eqnarray}
where $\omega_{1}=2a/L_{1}-1$, $\omega_{2}=2b/L_{2}-1$, $L_{1}=a+c$, and $L_{2}=b+c$.  For fixed $L$, moving the delta functions about in the 2-wire changes $a,b,c$ but not the total length, and thus the heights remain fixed.

The secular function for this graph is then easily seen to be
\begin{eqnarray}\label{secTwoDelta}
&& F_{2\delta}(g_{1},g_{2};a,b,c) = \\
&& F_{\delta}(g_{1};a,c)F_{\delta}(g_{2};b,c)-\sin{ka}\sin{kb}\nonumber
\end{eqnarray}
Setting $F_{2\delta}=0$ determines the eigenvalues of the graph.

A useful form of the secular function for the graph is found by multiplying out the factors in Eq. (\ref{secTwoDelta}) and rearranging the various terms.  This leads to
\begin{eqnarray}\label{secTwoDeltaExpanded}
&&\tilde{F}_{2\delta}(g_{1},g_{2};a,b,c)=\frac{4g_{1}g_{2}\sin{ka}\sin{kb}\sin{kc}}{(kL)^{2}}\nonumber \\
&+& \frac{2}{kL}\left(g_{1}\sin{ka}\sin{kL_{2}}+g_{2}\sin{kb}\sin{kL_{1}}\right)\nonumber \\
&+& \sin{kL}
\end{eqnarray}
The edge functions will have the canonical form, for nonvanishing amplitudes at the $\delta$ potential positions,
\begin{eqnarray}\label{edges}
\phi_{a}(s_{a}) &=& \frac{A\sin{ks_{a}}}{\sin{ka}} \\
\phi_{c}(s_{c}) &=& \frac{A\sin{k(c-s_{c})}+B\sin{ks_{s}}}{\sin{kc}}\nonumber \\
\phi_{b}(s_{b}) &=& \frac{B\sin{k(s_{b}-b)}}{\sin{kb}}\nonumber
\end{eqnarray}
where now we measure each edge from its origin at the start of each edge. The relative amplitudes of the edge functions in Eq. (\ref{edges}) may be obtained from Eq. (\ref{secTwoDelta}).

Replacing $k\rightarrow\imath\kappa$ in Eq. (\ref{secTwoDeltaExpanded}) leads to
\begin{eqnarray}\label{secTwoDeltaNeg}
&&-\imath \tilde{F}_{2\delta}(g_{1},g_{2};a,b,c)=\frac{4g_{1}g_{2}\sinh{\kappa a}\sinh{\kappa b}\sinh{\kappa c}}{(\kappa L)^{2}}\nonumber \\
&+& \frac{2}{\kappa L}\left(g_{1}\sinh{\kappa a}\sinh{\kappa L_{2}}+g_{2}\sinh{\kappa b}\sinh{\kappa L_{1}}\right)\nonumber \\
&+& \sinh{\kappa L}
\end{eqnarray}

The right hand side is zero at $\kappa L=0$ and is positive for $g_{1}>0$ and $g_{2}>0$, so there are no negative energy solutions for delta barriers.  If either or both of the delta strengths are negative, negative solutions are possible.  For large $\kappa L$, the right hand side is always positive.  Expanding the right hand side as a series in $\kappa L$ yields

\begin{equation}\label{secTwoDeltaNegApprox}
-\imath \tilde{F}_{2delta}\approx C_1(\kappa L)+C_3\frac{(\kappa L)^3}{3!}+C_5\frac{(\kappa L)^5}{5!}
\end{equation},
where

\begin{equation}\label{C1}
C_1=4g_{1}g_{2}\frac{abc}{L^{3}}+2g_{1}\frac{aL_{2}}{L^{2}}+2g_{2}\frac{bL_{1}}{L^{2}}+1,
\end{equation}

\begin{eqnarray}\label{C3}
C_3 &=& 4g_{1}g_{2}\frac{abc(a^2+b^2+c^2)}{L^5}\nonumber \\
&+& \frac{2g_1(aL_{2}^{3}+a^{3}L_{2})}{L^{4}}+ \frac{2g_{2}(bL_{1}^{3}+b^{3}L_{1})}{L^{4}}\nonumber \\
&+& 1
\end{eqnarray}

and

\begin{eqnarray}\label{C5}
C_5 &=& 4g_{1}g_{2}\frac{abc\left[a^4+b^4+c^4+\frac{5!}{3!3!}(a^2b^2+a^2c^2+b^2c^2)\right]}{L^{7}}\nonumber \\
&+& 2g_1\frac{aL_{2}^{5}+a^{5}L_{2}+\frac{5!}{3!3!}a^{3}L_{2}^{3}}{L^{6}}\nonumber \\
&+& 2g_2\frac{bL_{1}^{5}+b^{5}L_{1}+\frac{5!}{3!3!}b^{3}L_{1}^{3}}{L^{6}}\nonumber \\
&+& 1.
\end{eqnarray}

For a given set of edge lengths, there will be a range of g's for which $C_1<0$.  If $g_{2}>0$, then there will be a single negative energy solution when
\begin{equation}\label{g1NEG}
g_{1}<-\frac{1}{2}\left(\frac{L^{3}+2g_{2}bLL_{1}}{aLL_{2}+2g_{2}abc}\right)
\end{equation}
and vice-versa for $g_{1}\leftrightarrow g_{2}$.  A close examination of Eq. (\ref{secTwoDeltaNegApprox}) shows that this function can actually change sign twice for negative values of both g's.  This means there may be two negative energy solutions whenever both delta functions are wells.  Explicit forms for small values of $\kappa L$ may be found by setting $-\imath F_{2\delta} = 0$, which is a quadratic equation in $(\kappa L)^{2}$.

\begin{figure}\center
\includegraphics[width=4.5in]{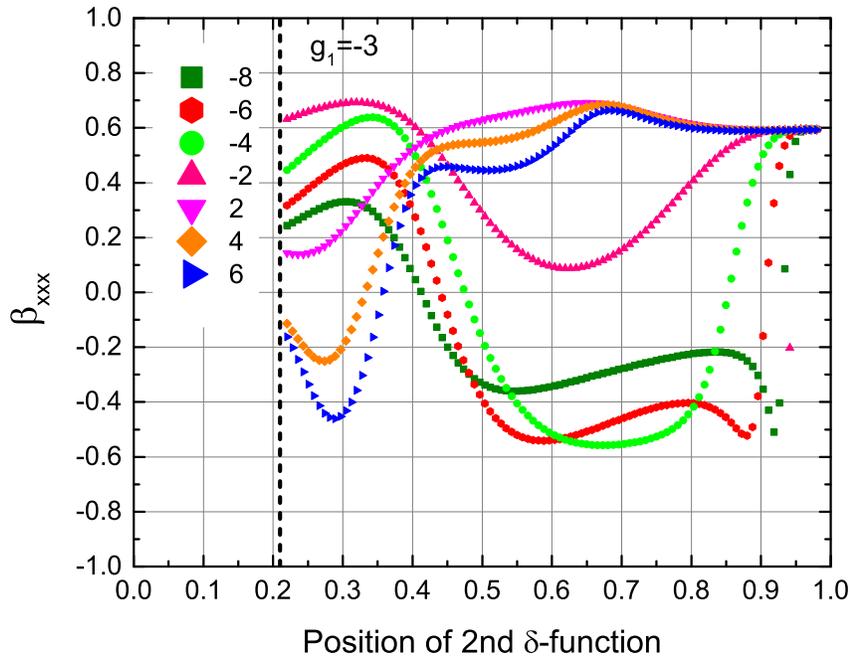}
\caption{Variation of $\beta_{xxx}$ with the position of the second $\delta$ function when the first is fixed.}\label{fig:2deltaBetaLines_g1=-3}
\end{figure}

\begin{figure}\center
\includegraphics[width=4.5in]{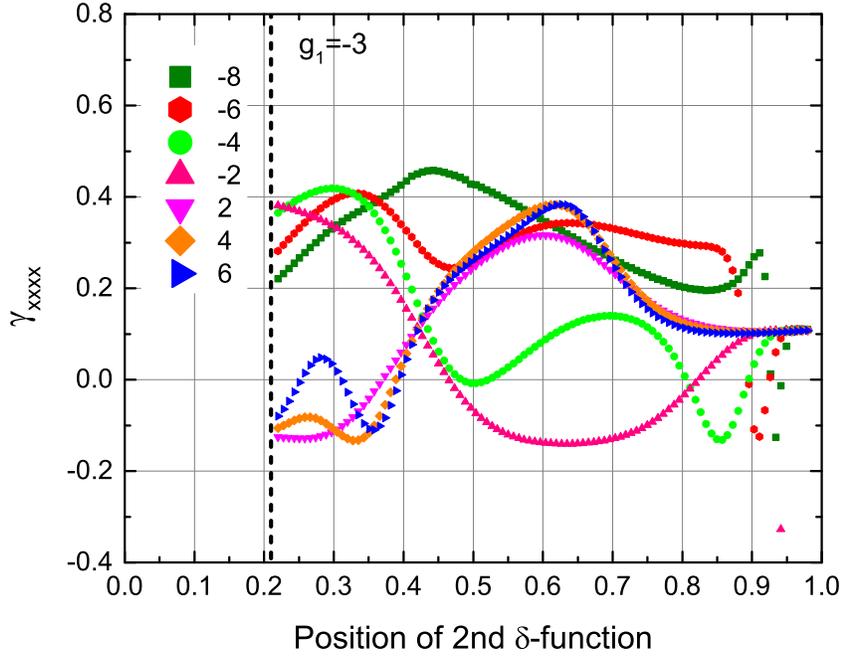}
\caption{Variation of $\gamma_{xxxx}$ with the position of the second $\delta$ function when the first is fixed.}\label{fig:2deltaGammaLines_g1=-3}
\end{figure}

Figures \ref{fig:2deltaBetaLines_g1=-3} and \ref{fig:2deltaGammaLines_g1=-3} shows the variation of $\beta_{xxx}$ (top)  and $\gamma_{xxxx}$ (bottom) with the position of the second $\delta$ function on the wire when the first is located at $a=0.21$.  The curves show the variation as $g_{2}$ ranges over various values, while $g_{1}=-3$.

The middle left panel of Figure \ref{fig:123delta_states_bestBetaGamma} displays the first seven eigenstates for the graph with the largest $\beta=0.697$.  The strengths of the two $\delta$ functions are shown in the panel.  All are negative.  The middle right panel of Figure \ref{fig:123delta_states_bestBetaGamma} displays the first seven eigenstates for the graph with the largest $\gamma=0.588$.  The strengths of the two $\delta$ functions are also shown in the panel.

\section{Compressed delta atom with three potentials}\label{sec:delta3Atom}

\begin{figure}\center
\includegraphics[width=3.4in]{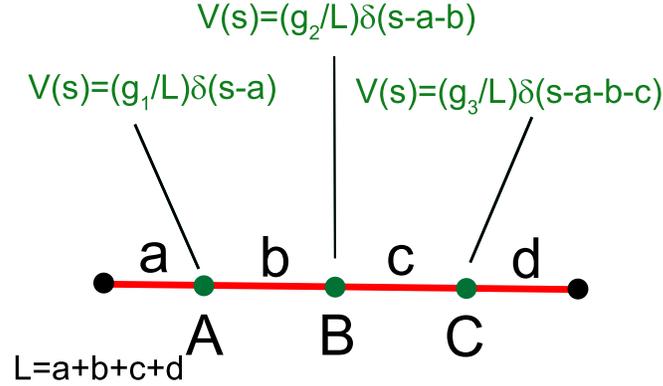}
\caption{Linear wire graph with three finite $\delta$ functions is comprised of three 2-wire delta motifs.}\label{fig:delta3delta}
\end{figure}

Consider now the use of the delta motif to create a graph with three delta potentials on a wire, as in Fig \ref{fig:delta3delta}.  Using the secular expression for the motif in Eq. (\ref{secReln}), we may immediately write down the coupled amplitude equations as
\begin{eqnarray}\label{ampEqns2}
A F_{\delta}(g_{1};a,b) &=& B\sin{ka} \\
B F_{\delta}(g_{2};b,c) &=& A\sin{kc}+C\sin{kb}\nonumber \\
C F_{\delta}(g_{3};c,d) &=& B\sin{kd}\nonumber
\end{eqnarray}
from which the secular function for this graph is easily seen to be
\begin{eqnarray}\label{secThreeDelta}
&& F_{3\delta}(g_{1},g_{2},g_{3};a,b,c,d) = \\
&& F_{\delta}(g_{1};a,b)F_{\delta}(g_{2};b,c)F_{\delta}(g_{3};c,d)\nonumber \\
&-& F_{\delta}(g_{1};a,b)\sin{kb}\sin{kd}-F_{\delta}(g_{3};c,d)\sin{ka}\sin{kc}\nonumber
\end{eqnarray}

Setting $F_{3\delta}=0$ determines the eigenvalues of the graph.  As might be anticipated, the 3-delta graph can have three negative energy  states with all three couplings are negative, two such states when two couplings are negative, or one when one coupling is negative.  For completeness, we show the secular equation in a form similar to those described for the one- and two-$\delta$ wires in Eq. (\ref{secThreeDeltaExpanded}).

\begin{eqnarray}\label{secThreeDeltaExpanded}
&&F_{3\delta}(g_{1},g_{2},g_{3};a,b,c,d) \\
&=& 8g_{1}g_{2}g_{3}\sin{ka}\sin{kb}\sin{kc}\sin{kd}\nonumber \\
&+& 4(kL)g_{1}g_{2}\sin{ka}\sin{kb}\sin{kL_{3}}\nonumber \\
&+& 4(kL)g_{2}g_{3}\sin{kc}\sin{kd}\sin{kL_{1}}\nonumber \\
&+& 4(kL)g_{1}g_{3}\sin{ka}\sin{kd}\sin{kL_{2}}\nonumber \\
&+& 2(kL)^{2}g_{1}\sin{ka}\sin{k(b+c+d)}\nonumber \\
&+& 2(kL)^{2}g_{3}\sin{kd}\sin{k(a+b+c)}\nonumber \\
&+& 2(kL)^{2}g_{2}\sin{kL_{1}}\sin{kL_{3}}\nonumber \\
&+& (kL)^{3}\sin{k(a+b+c+d)}\nonumber
\end{eqnarray}

The bottom left panel of Figure \ref{fig:123delta_states_bestBetaGamma} displays the first seven eigenstates for the graph with the largest $\beta=0.709$.  The strengths of the three $\delta$ functions are shown in the plot.  All are negative.  The bottom right panel of Figure \ref{fig:123delta_states_bestBetaGamma} displays the first seven eigenstates for the graph with the largest $\gamma=0.590$.  The strengths of the three $\delta$ functions are shown in the plot.

\begin{figure}\center
\includegraphics[width=5in]{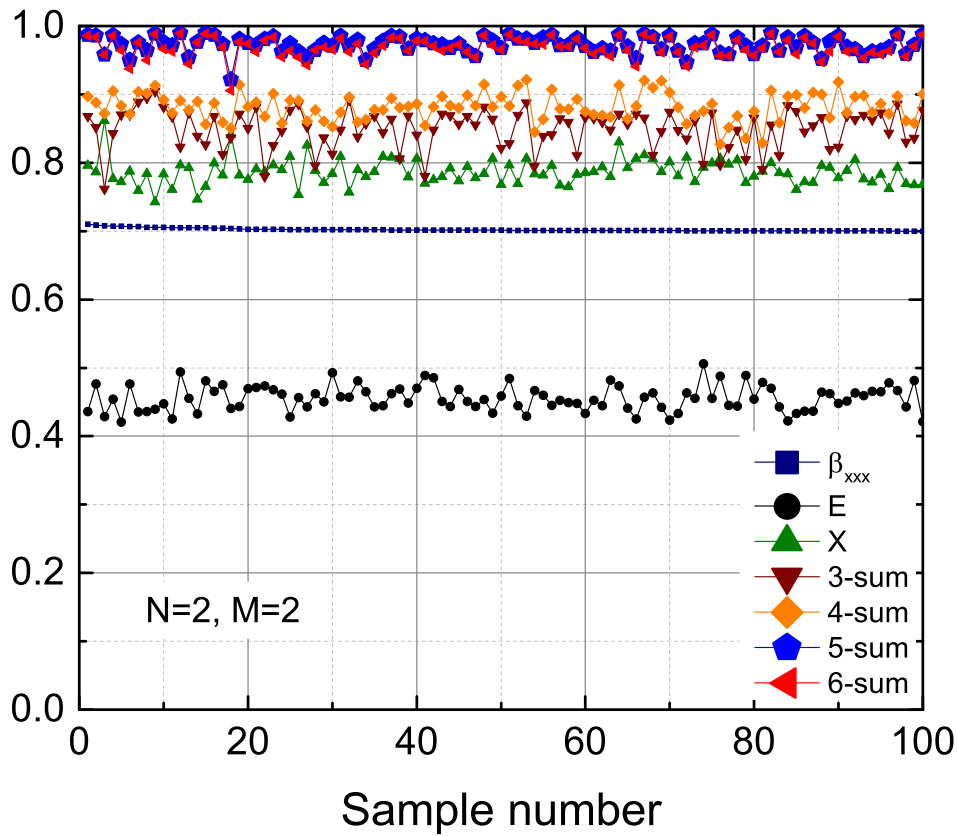}
\caption{Requirement for more than three states in the sum rules when $\beta_{xxx}$ is at or near the potential optimization limit.  In the units used in this paper, the diagonal sum rules always sum to unity. The figure shows that at least five states are required in the sum to (nearly) satisfy the sum rule.}\label{fig:3delta_EXbetaSums}
\end{figure}

Figure \ref{fig:3delta_EXbetaSums} illustrates how the three-level sum rule fails to hold for this graph when $\beta_{xxx}$ is at or very near its maximum.  In fact, five levels appear to be required to get the sum rule within a percent of its full value.

\begin{figure}\center
\includegraphics[width=5in]{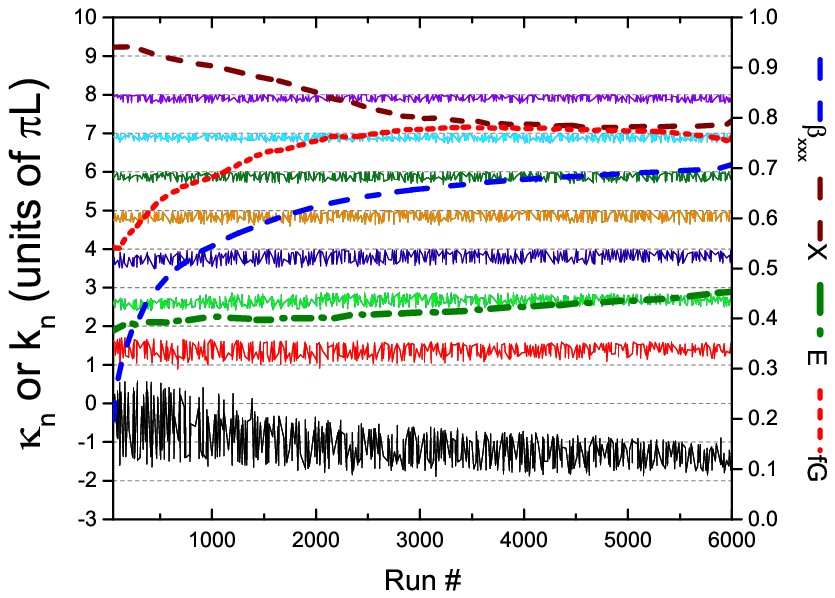}
\caption{Plot of the first eight eigenstates and of $\beta$ and the three-level model parameters X,E, and fG for a Monte Carlo run of $6000$ graphs with three $\delta$ functions with random strengths and random locations.}\label{fig:3delta_betaXEfG_kn}
\end{figure}

Fig. \ref{fig:3delta_betaXEfG_kn} is a snapshot view of a Monte Carlo run with $6000$ graphs having random sets of $\delta$ strengths at random locations.  The three-level parameter $X$ approaches the expected value of $0.79$ as $\beta\rightarrow 0.71$, but the energy ratio $E$ scales approaches $0.45$ as $\beta\rightarrow 0.71$, still short of the universal value of $0.5$ achieved for all prior potential optimization studies.

\begin{figure}\center
\includegraphics[width=5in]{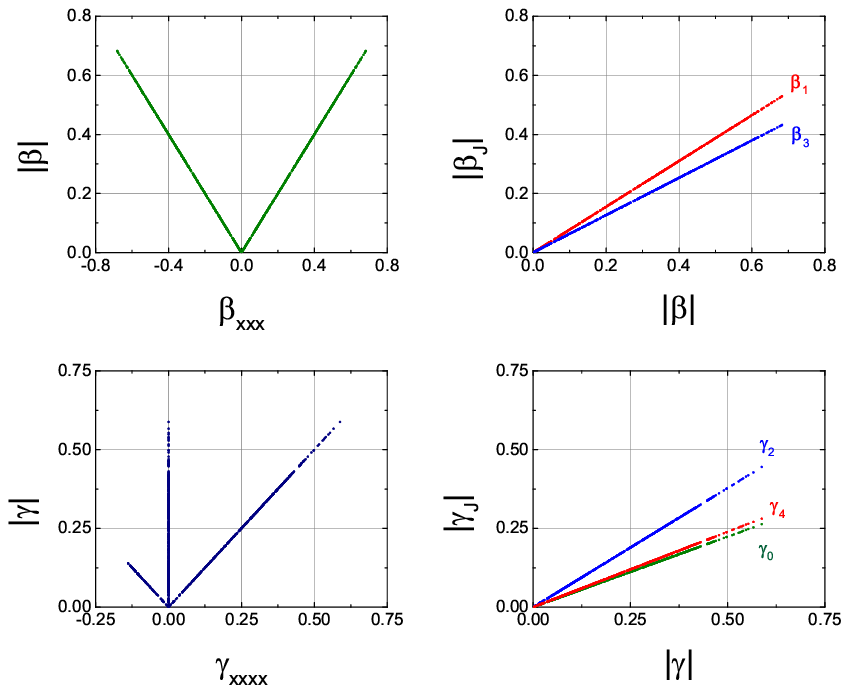}
\caption{Plot of the tensor norms and spherical tensor components for a Monte Carlo run of $10000$ graphs with three $\delta$ functions with random strengths and random locations. The vector spherical tensor component dominates $\beta{xxx}$ as expected for a straight wire.  The range of the tensor norms are exactly equal to those of the theory of fundamental limits.}\label{fig:3deltaTensorsStraightWires}
\end{figure}

Figure \ref{fig:3deltaTensorsStraightWires} illustrates the norms and spherical tensor components for these graphs.  The tensor norms are found to be bounded by the same limits as the diagonal component.

\section{Stars with a $\delta$ function at the central vertex}\label{sec:deltaStar}

Connected composite graphs may be constructed from the elemental graphs, or \emph{motifs}.  The spectra of connected graphs are the solutions to their secular equations, which always take the form of combinations of the secular functions of simpler graphs.  The N-star and lollipop motifs are sufficient to compute the states and spectra for all graphs \cite{lytel13.02}.

The nonlinearities of the N-star graph with edges terminated at infinite potential and the lollipop with its stick terminated at infinite potential have been calculated in the elementary QG model \cite{lytel13.01}. As isolated models of nonlinear, quantum confined systems, these are interesting structures because both topologies have intrinsic nonlinearities over three quarters of the fundamental limits from potential optimization.  Here, we are interested in dressing these graphs with a finite delta function at their central vertex to determine how their response changes and to examine their scaling properties.

For the 3-star with edges $a,b,c$, the secular function $F_{star}(a,b,c)$ for the graph is  \cite{pasto09.01,lytel13.01}
\begin{eqnarray}\label{3starSecularF}
F_{star}(a,b,c) &=& \frac{1}{4}\left[\cos{k_nL_1} + \cos{k_nL_2}\right. \nonumber \\
&+& \left.\cos{k_nL_3} - 3\cos{k_nL}\right],
\end{eqnarray}
where $L=a+b+c$, $L_1=|a+b-c|$, $L_2=|a-b+c|$, and $L_3=|a-b-c|$.

\begin{figure}\center
\includegraphics[width=3.4in]{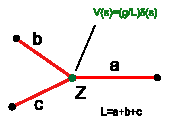}
\caption{Star graph with a delta function at the central vertex.}\label{fig:3star1Pdelta}
\end{figure}

The solutions to the secular equation $F_{star}=0$ for both rationally- and irrationally-related lengths have been discussed at length in Ref. \cite{lytel13.02}.  The energy eigenvalues are located in cells between root boundaries \cite{pasto09.01,lytel13.01}.  Their values move around within the root boundaries but the root boundaries are fixed and scale linearly with state number.

Consider now the dressed star graph.  The edge functions match at the central vertex, but the presence of the delta function changes the flux conservation condition \cite{kuchm04.01,kostr99.01,exner89.01} from its canonical form \cite{pasto09.01,lytel13.01} to

\begin{equation}\label{3starDeltaSecular}
F_{star}(a,b,c) +(2g/kL)\sin{ka}\sin{kb}\sin{kc}=0 .
\end{equation}

The solutions to Eq. (\ref{3starDeltaSecular}) for positive energies (real k) may be found numerically.  The root boundaries for nonzero $g$ are irregular and depend on the values of the edges, unlike the free star where the root boundaries are at multiples of $\pi/L$.  As with the compressed delta atom graphs, the delta star will support negative energy solutions when $g$ is negative and less than a critical value $g_c$ given by
\begin{equation}\label{gCrit}
g_{c}=-\frac{(a+b+c)(ab+ac+bc)}{2abc}.
\end{equation}
This may be seen by replacing $k\rightarrow\imath\kappa$, with real $\kappa$, leading to the negative energy secular equation
\begin{eqnarray}\label{3starSecularFnegE}
&& \frac{1}{4}\left[\cosh{\kappa L_1} +\cosh{\kappa L_2} +\cosh{\kappa L_3} -3\cosh{\kappa L}\right]\nonumber \\
&-& (2g/\kappa L)\sinh{\kappa a}\sinh{\kappa b}\sinh{\kappa c} = 0 .
\end{eqnarray}

\begin{figure}\center
\includegraphics[width=5.0in]{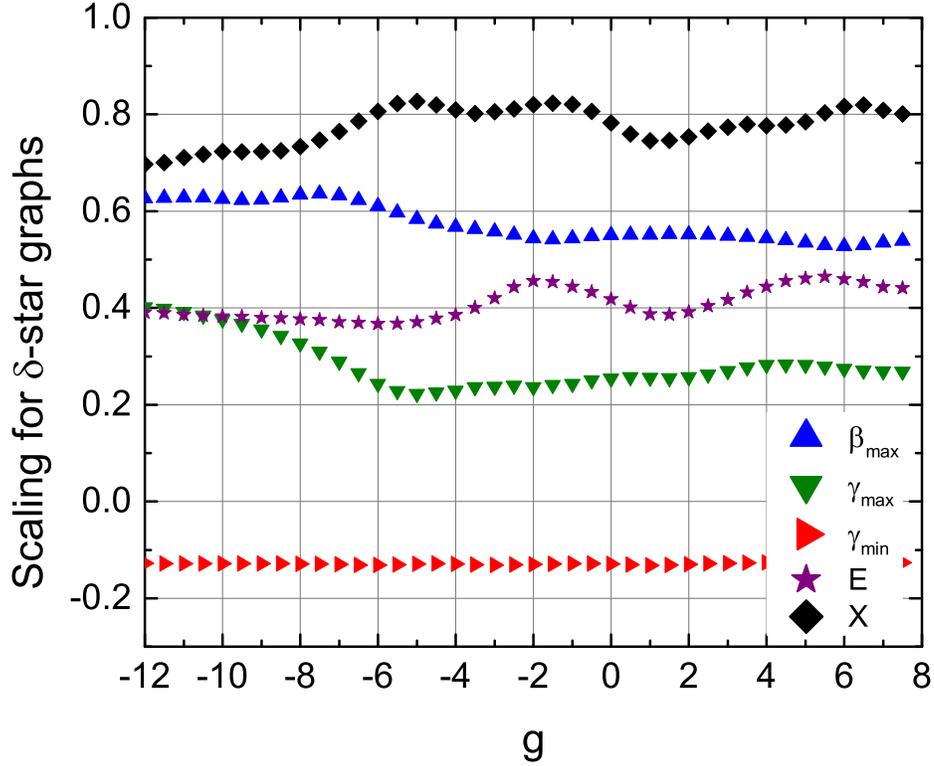}\center
\caption{Variation of $\beta_{xxx}$, $\gamma_{xxxx}$, $X$, and $E$ with the strength $g$ of the central potential.  The values for the bare star graph are those obtained at $g=0$. Star graphs have a range of hyperpolarizabilities determined by the prong lengths and the angles between them.  In this figure, the data points for both $\beta$ and $\gamma$ are indicated for their extreme values for any possible geometries of the graph with a given g, as indicated by the use of the subscripts max and min in the legend.}\label{fig:star1delta_beta_gamma_E_X_vs_g}
\end{figure}

We solve for the hyperpolarizabilities of the graphs in the usual way, computing the eigenvalues and transition moments, and then performing the sum over states in a Monte Carlo calculation, varying the edge lengths and angles with respect to a fixed external axis at various values of the delta strength $g$.  Figure \ref{fig:star1delta_beta_gamma_E_X_vs_g} displays the dependence of $\beta$ and $\gamma$ on the strength of the potential, and also displays the variation of the three level model parameters $X$ and $E$ as functions of $g$.

\section{Lollipop with a $\delta$ potential at the central vertex}\label{sec:deltaLollipop}

\begin{figure}\center
\includegraphics[width=3.4in]{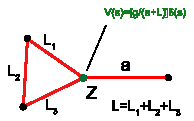}
\caption{Lollipop graph with a delta function at the central vertex.}\label{fig:lollipop1Pdelta}
\end{figure}

Consider now a lollipop graph with a delta function at its central vertex, as shown in Fig \ref{fig:lollipop1Pdelta}.  We have normalized the potential so that $g$ is dimensionless and that the length scale is $a+L$ so that any scaling of length for the entire graph scales the potential the same way.  The secular function for the lollipop without the delta function has been previously derived \cite{lytel13.01} and is
\begin{equation}\label{lollipopSecularF}
F_{pop}(a,L) = \frac{1}{2}\left[3\cos{k(a+L/2)} - \cos{k(a-L/2)}\right] ,
\end{equation}
where $L=L_1+L_2+L_3$ is the loop length.  The secular equation $F_{pop}=0$ generates the eigenvalues for the graph.

The wavefunctions of the lollipop graph are a composite of two sets of wavefunctions, one set that is nonzero at the central vertex and on all edges, and one for wavefunctions that vanish at the origin and are exactly zero on the prong edge.  The first set correspond to the symmetric wavefunctions of a 3-sided bent wire (open at the central vertex) coupled to a nonzero prong wavefunction, while the second set correspond to the asymmetric wavefunctions of a 3-sided bent wire (open at the central vertex) with a zero prong wavefunction.  When another graph is attached to the prong, the loop-only wave functions go away and we're left with the wave functions satisfying the secular equation above.

With the delta function, the secular equation is modified to
\begin{equation}\label{lollipopDeltaSecular}
F_{pop}(a,L) +\frac{2g}{k(a+L)}\sin{ka}\cos{kL/2}=0.
\end{equation}
As with the star graph having a delta function at its central vertex, the lollipop with the delta function at its central vertex may also have negative energy solutions for certain ranges of the strength $g$.  Setting $k\rightarrow\imath\kappa$ converts Eq. (\ref{lollipopDeltaSecular}) to
\begin{eqnarray}\label{lollipopDeltaSecNeg}
&& \frac{1}{2}\left[3\cosh{k(a+L/2)} - \cosh{k(a-L/2)}\right]\nonumber \\
&+& \frac{2g}{k(a+L)}\sinh{ka}\cosh{kL/2}\nonumber \\
&=& 0.
\end{eqnarray}

\begin{figure}\center
\includegraphics[width=5.0in]{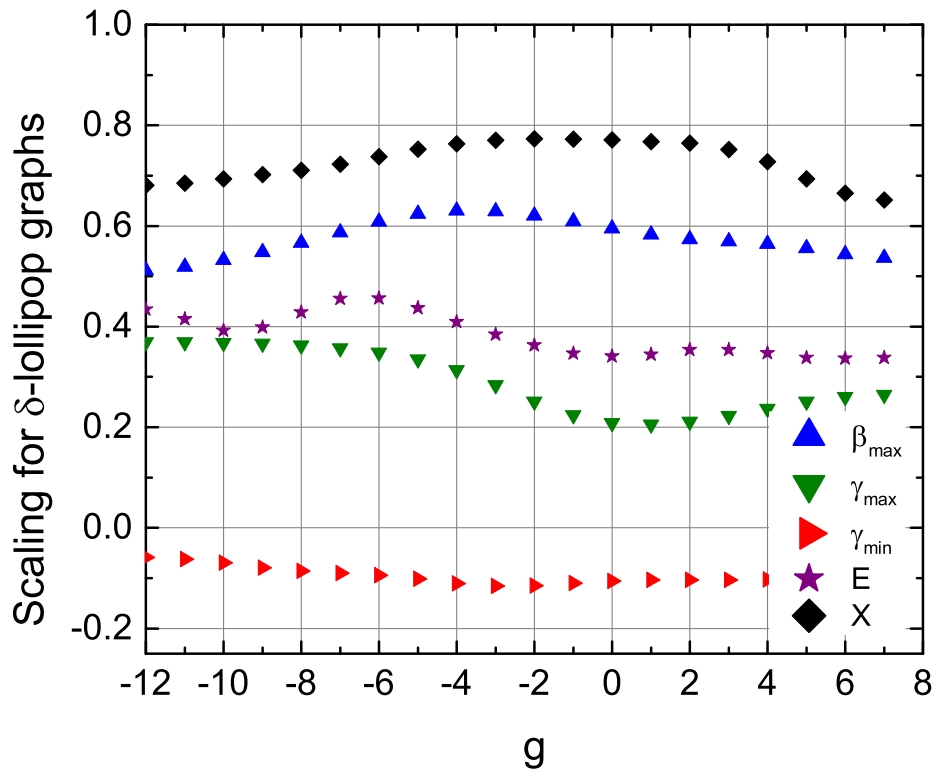}\center
\caption{Variation of $\beta_{xxx}$, $\gamma_{xxxx}$, $X$, and $E$ with the strength $g$ of the central potential..}\label{fig:lollipop1delta_beta_gamma_E_X_vs_g}
\end{figure}

We solve for the hyperpolarizabilities of the graphs in the usual way, computing the eigenvalues and transition moments, and then performing the sum over states in a Monte Carlo calculation, varying the edge lengths and angles with respect to a fixed external axis at various values of the delta strength $g$.  Figure \ref{fig:lollipop1delta_beta_gamma_E_X_vs_g} displays the dependence of $\beta$ and $\gamma$ on the strength of the potential, and also displays the variation of the three level model parameters $X$ and $E$ as functions of $g$.

\section{Conclusions}\label{sec:Conclusions}

The fundamental limits of the first and second hyperpolarizabilities of nonlinear optical structures first showed that all measurements to date of real materials fall well short of the limits \cite{kuzyk00.01,kuzyk00.02}.  Extensive analytical studies using Monte Carlo methods that employ sum rules to constrain the space of energies and transition moments have shown that the limits are exactly achieved.  No known system is able to generate the maxima, though exotic Hamiltonians are expected to be able to achieve this.  Invoking additional constraints on the energies by using specific potential functions has shown that the fundamental limits decrease by about thirty percent.  Prior to the work presented in this paper, no analytical model that begins with a potential and calculates the energies and transition moments has generated the potential optimization results.

This paper is the first to present such a model, dressed quasi-one dimensional elementary quantum graphs.  Our prior work on the relationship between the topology of a bare (undressed) graph and the range of its nonlinearities showed that the best achievable results were under three quarters of the potential optimization limits.  That result revealed that both the two fundamental motifs, the star and the lollipop, were the best topologies for which select geometries could be found that had these responses.  In contrast, the bare straight or bent wire had a much smaller response, regardless of its shape.  The three star graph is topologically distinct from a wire.  But the change in boundary conditions due to the addition of a prong to the wire creates a discontinuity in the edge function slopes that is similar to that obtained by inserting a $\delta$ function in a wire segment.  This suggested that dressing the wire graph with a $\delta$ function would significantly increase the response of the graph.

The work in this paper showed that a wire dressed with a single, finite $\delta$ potential well produces a response nearly at the fundamental limits from potential optimization.  Adding another $\delta$ function increases the response to the fundamental limits.  Adding a single $\delta$ potential at the vertex of the star or lollipop graph likewise increases the response to approach the fundamental limits.  Interestingly, in this case, the addition of the well increases the response but not as much as in a wire due to the constraints imposed by the multiple edges in the star or lollipop.  A single wire with two $\delta$ functions offers sensitive tuning of the localization of the eigenstates to produce the best possible overlap for maximum response.

Studies of dressed quantum graphs have ramifications in applications and in enhancing our fundamental understanding of how geometry, topology, and control of the wavefunctions using dressing affect the interaction between light and matter.  For example, delta potentials are to a good approximation a defect in a quantum wire.  As such, it may be straightforward to intentionally place a defect in a quantum wire to greatly enhance its nonlinear optical response.  Composite materials filled with such quantum wires could lead to materials with ultra-large nonlinear susceptibilities.  Furthermore, the tuning of the energy spectrum afforded by the strength of the defect might we usable to control the spectral response of the material to meet the requirements of a particular application.

The precise control of a quantum system through the addition of defects allows a broader sampling of the configuration space to which all real systems are confined.  One nagging question is the 30\% gap between the limit and the largest values obtained from theoretical calculations of a large number of quantum systems.  The key may lie in the three-level ansatz, which was used to derive the limits. Empirical evidence supports the fact that the hyperpolarizability is always well approximated by a three-level system near a local maximum, but why has the global maximum predicted by the limits eluded all calculations and measurements?

The functions defined by Equations \ref{eff} and \ref{Gee} are assumed to be independent in the calculation of the limits, but in real systems, they may be related.  As depicted in Figures \ref{fig:1delta_spectra_beta_VS_g} and \ref{fig:123delta_states_bestBetaGamma}, as the negative delta function is made stronger, the energy gap increases, yielding a favorable energy spectrum.  However, as the delta function gets stronger, the ground-state wavefunction becomes more localized, so the overlap with other states is reduced, thus decreasing the nonlinearity.  As such, the parameters $E$ and $X$ in real systems may not be independent.  If this relationship can be quantified in the form of a constraint, it would undoubtedly be a useful tool to understand the puzzling observations.

The various patterns observed in the present studies hint at a rich underlying structure that is a manifestation of the constraints imposed by nature.  The present study is a first step in accumulating the needed evidence to unravel the mystery that presents itself.  Future studies are focusing on developing general principles that are suggested from the present observations.

\section{Acknowledgments}
We would like to thank the National Science Foundation (NSF) (ECCS-1128076).

\end{document}